\documentclass{aa}  

\usepackage{graphicx}
\usepackage[normalem]{ulem}
\usepackage{txfonts}
\usepackage{xcolor}
\usepackage{bm}

\newcommand{\subsubsubsection}[1]{\paragraph{#1}\mbox{}\\}
\newcommand{\ergs}{\,ergs\,s$^{-1}$} 
\newcommand{\kms}{\,km\,s$^{-1}$} 
\newcommand{\uJy}{\,$\mu$Jy\,bm$^{-1}$} 
\newcommand{\solarmass}{\(\textup{M}_\odot\)}

\begin{document} 

   \title{Quantifying jet--interstellar medium interactions in Cyg X-1: Insights from dual-frequency bow shock detection with MeerKAT}
   \titlerunning{bow shock structure due to Cyg\,X-1 jets}
   \author{P. Atri
          \inst{1,} \inst{2},
          S. E. Motta\inst{3,} \inst{4},
          Jakob van den Eijnden
          \inst{5,6},
          James H. Matthews
          \inst{4},
          James C.A. Miller-Jones
          \inst{7},
          Rob Fender
          \inst{4},
          David Williams-Baldwin,
          \inst{8}
           Ian Heywood
          \inst{4,9,10} and
          Patrick Woudt
          \inst{11}
          }
    \authorrunning{Atri and Motta et al. }
   \institute{ASTRON, Netherlands Institute for Radio Astronomy, Oude Hoogeveensedijk 4, 7991 PD Dwingeloo, The Netherlands\\
              \email{atri@astron.nl}
        \and 
        Department of Astrophysics/IMAPP, Radboud University, P.O. Box 9010, 6500 GL, Nijmegen, The Netherlands 
        \and 
        Istituto Nazionale di Astrofisica, Osservatorio Astronomico di Brera, via E.\,Bianchi 46, 23807 Merate (LC), Italy\\
            \email{sara.motta@inaf.it}
        \and 
        University of Oxford, Department of Physics, Astrophysics, Denys Wilkinson Building, Keble Road, OX1 3RH, Oxford, United Kingdom
        \and 
        Department of Physics, University of Warwick, Coventry CV4 7AL, UK
        \and
        Anton Pannekoek Institute for Astronomy, Universiteit van Amsterdam, Science Park 904, 1098, XH, Amsterdam, The Netherlands
        \and
        International Centre for Radio Astronomy Research, Curtin University, GPO Box U1987, Perth, WA 6845, Australia
        \and
         Jodrell Bank Centre for Astrophysics, School of Physics and Astronomy, The University of Manchester, Manchester, M13 9PL, United Kingdom
        \and 
         Department of Physics and Electronics, Rhodes University, P.O. Box 94, Makhanda, 6140, South Africa
         \and
         South African Radio Astronomy Observatory, 2 Fir Street, Observatory 7925, South Africa
         \and
         Department of Astronomy, University of Cape Town, Private Bag X3, 7701 Rondebosch, South Africa 
             \thanks{}
             }
   \date{Received xx yy, 2024; accepted abc xx, 2025}

 
  \abstract
   {Accretion and outflows are astrophysical phenomena observed across a wide range of objects, from white dwarfs to supermassive black holes. Developing a complete picture of these processes requires complementary studies across this full spectrum of jet-launching sources. Jet--interstellar medium (ISM) interaction sites near black hole X-ray binaries provide unique laboratories that provide insights into the energetics of the jets launched from stellar-mass black holes.}
   {This work aims to detect and characterise the bow shock near one black hole X-ray binary, Cyg\,X-1, and then use this bow shock structure to parametrise the properties of the jet launched by Cyg\,X-1 over its lifetime. }
   {We used the MeerKAT radio telescope to investigate the bow shock structure formed by the interaction between the jets of Cyg\,X-1 and the ISM. Using new L- and S-band detections of the bow shock, we constrained the density of the unshocked ISM and mapped the bow shock’s spectral index. These values were applied to self-similar models developed initially for FR{\sc ii} galaxies to estimate the energy transport rate and the age of Cyg\,X-1 jets.
   }
   {We successfully detect the bow shock north of Cyg\,X-1 in the L and S bands and report its size and brightness. We present the spectral index distribution across the bow shock, which is in the range -0.9$\leq \alpha \leq$ 0.4, with an error distribution (0.6$\leq \Delta\alpha \leq$ 1.5) that peaks at unity. We determine that the unshocked ISM density is 6--7\,cm$^{-3}$ for a temperature range of 10$^4$--3$\times$10$^6$\,K. This temperature range suggests that the velocity of the bow shock is $21$\,km$\,$s$^{-1}<\dot{L} <\,364\,$km\,s$^{-1}$. The age of the Cyg\,X-1 jet responsible for the bow shock is 0.04 -- 0.3\,Myr, and the power of the jet is constrained to 2$\times10^{31}$\,ergs s$^{-1}<$ Q$_{jet}^{a}<$ 1$\times10^{35}$\,ergs s$^{-1}$ for the case of opening angles of 0.3$^\circ$--2.0$^\circ$. We also detect new morphological features of the bow shock in the S-band image. The comparison of archival H$_\alpha$ maps with the new radio observations hints at different regions of emission, different temperature ranges, and different ISM densities.}
   {MeerKAT’s sensitivity and resolution effectively reveal low surface brightness features of the Cyg\,X-1 bow shock. The spectral index suggests a consistent emission origin across the structure. The ISM density around Cyg X-1 is on the higher end for Galactic environments, and our results indicate a lower jet energy transport rate than prior estimates. Further searches with MeerKAT will help build a statistically significant sample, advancing our understanding of black hole X-ray binary jets and their impact on their local environments. }


   \maketitle
%
\section{Introduction}\label{Introduction}
Supermassive black holes (SMBHs; $\geq 10^{5}$\,\solarmass) at the centres of some active galactic nuclei (AGNs) and stellar-mass black holes ($\leq 100$\,\solarmass) in binary systems are known to launch relativistic jets. These jets transport a large amount of energy from the nucleus back into the galaxy, thus coupling the growth of the galaxy to the jets \citep[e.g.][]{McNamara2005,Croton2006,Nesvadba2010,Krause2023}. Black hole X-ray binaries (BHXBs), binary systems of stellar-mass black holes in accreting systems with main sequence stars, are considered the lighter analogues of SMBHs. The impact of BHXB jets on the interstellar medium (ISM) is not easily observable, unlike the interactions of AGN jets with the intergalactic medium (IGM). This is due to the different environments that AGNs and BHXBs reside in \citep{Heinz2002}. Unlike AGNs, BHXBs do not reside in dark matter halos and are susceptible to natal kicks \citep{Repetto2012, Atri2019}. BHXBs can thus be found in the Galaxy with high space velocities and can constantly change their environment.
Additionally, the ISM provides a lower barrier to the high-ram-pressured jets from BHXBs, and this results in weaker shock fronts. However, studies have noted the effect of BHXB jets on their local and Galactic-scale surroundings. \citet{Heinz2008} concluded that the plasma released by BHXB jets can fill the disc and halo of our Galaxy. This magnetised plasma can introduce sufficient seed magnetic fields in the ISM to produce the average magnetic field of our Galaxy. The plasma from jets that mixes with the ISM also gives rise to cosmic rays in the ISM \citep{Alfaro2024}. Thus, accurate estimates of the energy injected into the ISM by the relativistic jets of BHXBs are essential for modelling the Galactic magnetic field and the contribution of BHXBs to Galactic cosmic rays.

Active galactic nuclei and BHXBs launch relativistic jets that facilitate the transfer of the accreted power back to the inter-cluster medium and ISM, respectively. 
One way to understand the relation between accreting matter and the mechanism by which this energy is injected back into the environment is to study the interaction of the radio jets with their surroundings. Estimating various jet properties like the jet speed, the total jet power, the composition of the jet, and the magnetic field at the base of the jet is a challenging task, influenced by a range of assumptions, and has been attempted using a few different techniques \citep[e.g.][]{Russell2008,Tetarenko2018,Fender2019}. The interaction sites and bow shocks around BHXBs are a probe into many of these jet properties \citep[e.g.][]{Burbidge1959,Heinz2006}. \par 

Over the decades, gradual improvements have been made in modelling how the jets of AGNs and BHXBs interact with the ISM. These interactions give rise to observable features like radio lobes, radio bow shocks, and large X-ray cavities in AGNs \citep[e.g.][]{Scheuer1974,McNamara2007}. Early models suggested that radio lobes and hot spots in FR{\sc ii} galaxies require continuous energy input, leading to the idea that jets inflate cocoons that drive bow shocks \citep{Longair1973, Blandford1974}. Relativistic jets launched from BHXBs have also been observed to create radio structures as they interact with the ISM \citep[e.g.][] {Gallo2005,Panferov2017}. The jets launched from a BHXB are expected to sweep up the gas that is in their path, creating a cavity as they propagate from their launch location. The ejected jet plasma can continue into the medium as long as the swept-up mass is comparable to the mass of the ejected plasma and the energy losses of the jet are minimal during propagation. In some cases, the ejecta can propagate farther from the black hole system even if these conditions are not met due to previously ejected material already tunnelling a channel in the ISM. The jet material ultimately reaches the end of this tunnel and interacts with the denser ISM. This creates inflated radio lobes or bubbles far from the base of the black hole jet, and these can lead to bow shocks that expand sideways and forwards from these radio lobes.  

Since BHXBs are often found in lower-density environments than AGNs in a dynamical sense, bright jet--ISM interactions require local density amplifications \citep{Heinz2002,Carotenuto2024}. For example, the nebular structure observed around SS433 is modelled as a combination of a dense supernova remnant and the jets from SS433 creating shocks when interacting with this high-density remnant \citep[e.g.][]{Begelman1980,Panferov2017}. More recently, a bow shock structure has been discovered near GRS\,1915+105 \citep{Motta2025} adjacent to a large H{\sc ii} region. The bow shock structure detected near Cyg\,X-1 is close to the tail of a dense H{\sc{ii}} region, the Tulip Nebula. The Tulip Nebula is considered to be the leading cause of an increase of the local ISM density that has made the jet--ISM interaction near Cyg\,X-1 detectable as a bow shock structure \citep{Gallo2005}.

Cygnus\,X-1, or Cyg\,X-1, is a high-mass BHXB at a distance of 2.2$\pm$0.2\,kpc consisting of a massive 21.2$\pm$2.2\,\solarmass \ black hole with a blue supergiant of mass 41$\pm$8\,\solarmass \ in a 5.6-day orbit \citep{Miller-Jones2021}. Cyg\,X-1 was first discovered in 1971 \citep{Hjellming1971} and has been the subject of detailed studies and observations owing to its bright nature. BHXBs are jet-dominated in their hard X-ray state, wherein the radio emission is thought to be driven by partially self-absorbed jets \citep{Fender2004}. For Cyg\,X-1, the hard X-ray state radio emission has been resolved as a collimated jet of size 30\,AU \citep{Stirling2001}. BHXBs also have an intermediate state, which is associated with the BHXB leaving the hard X-ray spectral state and moving towards the soft state. During this state, ejecta from the system that is are causally disconnected from the black hole are observable at radio wavelengths \citep{Fender2006}. The energy injected into the ISM by a BHXB is thus due to jet activity in both these states.  \par
\citet[G05 hereafter]{Gallo2005} observed Cyg\,X-1 with the Westerbork Synthesis Radio Telescope (WSRT) at 1.4\,GHz and discovered a shock bubble being blown by the jets of Cyg\,X-1. G05 detected a ring-shaped, diffuse gas structure north of Cyg\,X-1 with a diameter of 5\,pc. The ring was seen in a direction perpendicular to the direction of the motion of Cyg\,X-1; this, combined with the fact that Cyg\,X-1 is one of the slowest moving X-ray binaries (XRBs) in our Galaxy (9$\pm$2\,mas\,yr$^{-1}$), allows us to confidently rule out that the structure detected was a shock left in the wake of fast-moving objects. Additionally, the motion of Cyg\,X-1 cannot be traced back to the centre of the ring, ruling out that the structure could be the supernova remnant of the black hole in Cyg\,X-1. Later, optical spectroscopic studies of the bow shock structure also confirmed that the origin of the emission could not be related to a supernova remnant \citep{Sell2015}. The direction of the hard state jet of Cyg\,X-1 is consistent with the scenario that the jet is responsible for creating the bow shock \citep{Stirling2001}. Combined with H$_\alpha$ observations, it was clear that the ring was emitting Bremsstrahlung radiation, and this thermally emitting ring was estimated to be moving with a velocity of 100-360\kms. With a well-informed assumption that Cyg\,X-1 spends 90\% of its lifetime in a hard state \citep{Grinberg2013}, the total power of the jet averaged over the lifetime of Cyg\,X-1 lies in the range 9$\times$10$^{35}$ $\leq$ P$_{jet}$ $\leq$ 10$^{37}$ \ergs (G05). \par
The WSRT is an east-west array with a maximum baseline of 2.7\,km, and its resolution is twice as high in the east-west direction as compared to the north-south direction for the declination of Cyg\,X-1 (35$^\circ$). The map of the ring around Cyg\,X-1 in G05 has a resolution of 25$\times$14 arcsec$^{2},$ and the noise level around the ring is of the order of 70\uJy, for a total observing time of 60\,hours. A more precise measurement of the total jet power, the velocity of the ring, and the average energy input to the ISM requires higher-precision estimations of the observed parameters of the ring-like structure. The MeerKAT radio telescope provides the required resolution and sensitivity to improve the WSRT radio map of Cyg\,X-1 in the L band (1.4\,GHz). With a maximum baseline of 8\,km, a total of 64 dishes, and a condensed core, this array provides a great opportunity to image diffuse emission structures like that of the Cyg\,X-1 bow shock.
We obtained a deeper and higher-resolution map of the structure using the MeerKAT telescope in two frequencies, produced the first spectral index map of the bow shock structure, combined these observations with archival optical data to get more accurate estimates of the ISM density, and ultimately modelled the jet parameters of Cyg\,X-1 that could be responsible for creating the bow shock structure. \par
The paper is structured as follows: Section \ref{Section 2} contains the description of the observations that were carried out, Sect. \ref{Section 3} lays out the process followed to reduce the data, Sect. \ref{Section 4} presents the results and the image maps obtained, and Sect. \ref{Section 5} discusses the implications of the maps. Finally, we conclude this paper with Sect. \ref{Section 6}.
\section{Observations}\label{Section 2}
We observed the field of Cyg\,X-1 using MeerKAT in two frequency bands, the L band and the S band under the ThunderKAT and the XKAT programmes, respectively. The ThunderKAT programme was a Large Survey Program that conducted weekly observations of active transient fields for the years 2018-2023. The XKAT programme is a MeerKAT open call project, which focused solely on observing fields of active XRBs and ran for the years 2023-2024. The L-band observations were conducted on 16 April 2022 (observation block ID 1650088876) with a total bandwidth of 856\,MHz centred at a frequency of 1284\,MHz. The pointing centre of the observations was right ascension (RA, J2000) $19^{h}:58^{m}:21^{s}.67$ and declination (Decl, J2000) of $35^{d}:11^{m}:50^{s}.70$. The field was observed for 15 minutes and was bracketed by a 2-minute observation of the phase and amplitude calibrator, J2015+3710 (RA $20^{h}:15^{m}:28^{s}.73$ and Decl $37^{d}:10^{m}:59^{s}.50$). We also used J1939-6342 (RA $19^{h}:39^{m}:25^{s}.03$ and Decl $-63^{d}:42^{m}:45^{s}.60$) as a bandpass and flux calibrator, which was observed for 5 minutes at the beginning of the observing block.  

The S-band observations were carried out on 14 March 2024 (Observation block ID 1710398476) at a central frequency of 2625\,MHz with a total bandwidth of 875\,MHz. The pointing centre of the observations was RA $19^{h}:58^{m}:14^{s}.23$ and Decl $35^{d}:16^{m}:22^{s}.10$, and the field was observed for 30\,minutes. The target observations were bracketed by a 2-minute observation of a phase and amplitude calibrator, J2236+2828 (RA $22^{h}:36^{m}:22^{s}.47$ and Decl $28^{d}:28^{m}:57^{s}.40$). The bandpass and flux calibrator used were the same as for the L-band observations.

\section{Data reduction} \label{Section 3}
MeerKAT data observed as a part of the ThunderKAT and XKAT programmes are reduced using {\tt{oxkat}} \citep{Heywood2020}, a series of Python code written as wrappers around common radio astronomy software to facilitate a well-calibrated and quick image of the observed field. Depending on the observed field and requirements of the object of interest (e.g. compact or extended emission), further adjustments to the input parameters to these Python scripts are made (e.g. negative or positive Briggs weighting). For both L-band and S-band data reduction on the ILIFU/idia cluster, the process follows three stages.\par
The first stage (1GC.py) averages the measurement set (MS) into 1024 frequency channels and creates a new MS. This is followed by applying basic flagging, for example, to known bad frequency channels, and edges of the frequency bandpass. \texttt{flagdata}, an autoflagging algorithm in {\tt{CASA}} \citep{CASA2022} is also used to identify outliers and flag them in the calibrator data. Once the basic flagging is done, \texttt{setjy} is used to set the flux density scale using the flux calibrator (J1939-6342) and obtain the delay, bandpass and gain solutions. This model is applied to both the phase calibrator (J2015+3710 for the L band and J2236+2828 for the S band) and the target. The amplitude and phase solutions are then obtained using the nearby phase calibrator and applied to the target field (Cyg X-1), which is then split into a new corrected MS. \par
The second stage (FLAG.py) runs the \texttt{Tricolor} package, which is specifically written for MeerKAT to do automatic Radio Frequency Interference mitigation in the MSs. The first unconstrained image of the fields is then created using \texttt{wsclean}, in that no masks or models are used in creating the image. We used a multi-scale approach for the cleaning algorithm that used Briggs weighting of 0.3 as the best compromise between sensitivity and resolution. An initial cleaning mask that encapsulates most of the bright emission in the field is created using a local rms threshold, and the level of this local rms can be set as required (see the config.py file uploaded on GitHub for the exact input parameters)\footnote{https://github.com/pikkyatri/CygX-1-bow-shock}.\par
The third stage of the scripts (2GC.py) focuses on direction-independent self-calibration of the corrected visibilities from the FLAG.py stage. The initial mask produced by FLAG.py is used to do a masked deconvolution of the data. After this stage, \texttt{Cubical} \citep{Cubical2018} is used to perform self-calibration of the MS, which is followed by refining the mask using this self-calibrated data. Finally, the visibilities are modelled with this new, refined mask. To further improve calibration, we ran the 2GC.py two more times by using the final mask produced as the input mask instead of using the mask made by the FLAG.py script. The resulting calibrated image fits file was then assessed using the Cube Analysis and Rendering Tool for Astronomy \citep[CARTA; ][]{comrie2018}.

\section{Results} \label{Section 4}
\subsection{L band}
\begin{figure*}
\centering
\resizebox{\hsize}{!}{\includegraphics{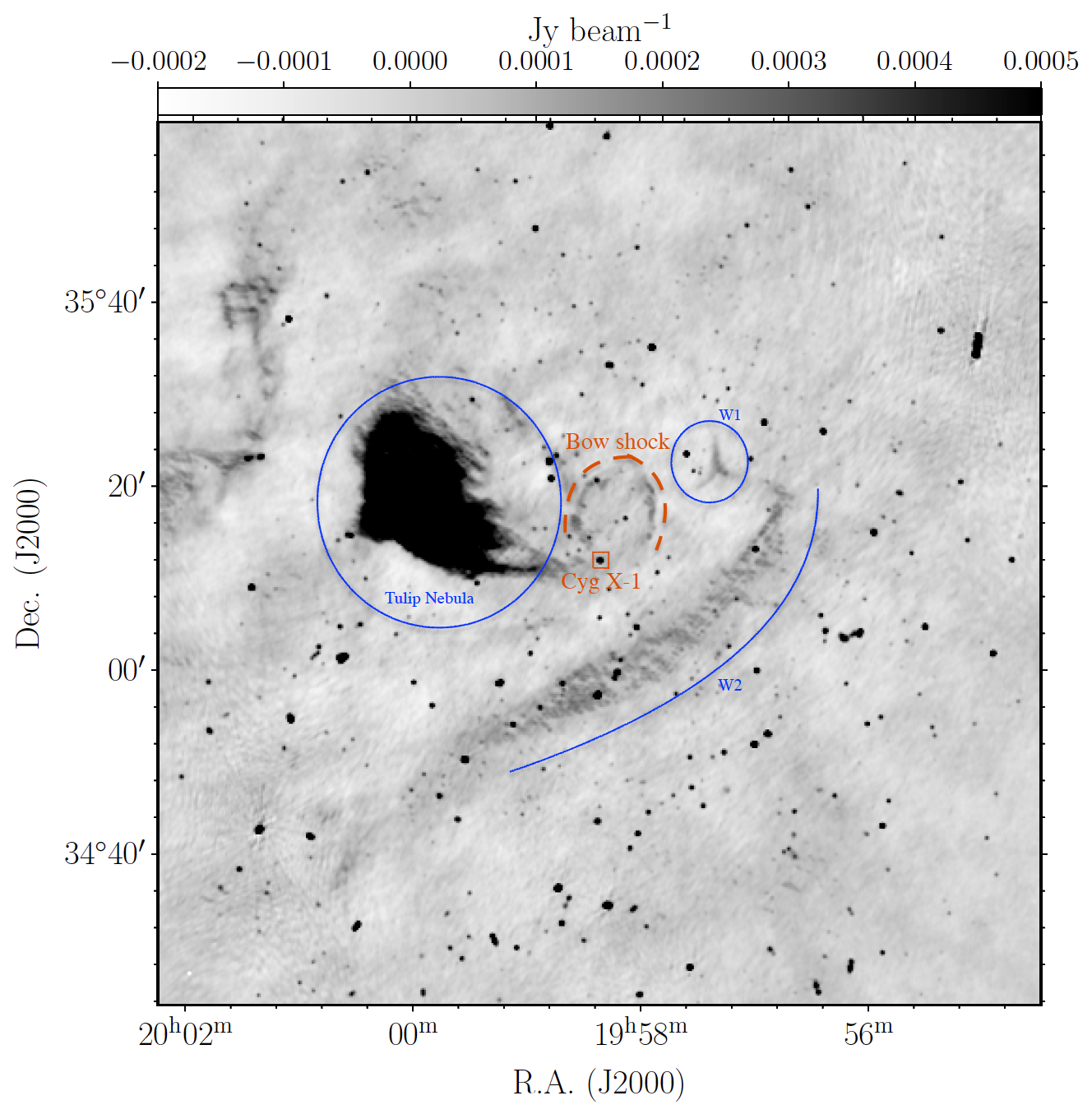}}
\caption{Cyg\,X-1 field as observed by the MeerKAT telescope in the L band. The orange square marks the location of Cyg\,X-1, which is seen as a compact source in the middle of the square. The dashed orange arc to the north of Cyg\,X-1 denotes the outer edge of the bow shock structure. The H{\sc{ii}} region known as the Tulip Nebula is seen on the eastern side of Cyg\,X-1. W1 and W2 are two new diffuse emission regions detected in the image to the north-west and south of Cyg\,X-1, respectively. }
\label{fig:full_map_Lband} 
\end{figure*}

\begin{figure*}
    \centering
    \begin{tabular}{ccc}
    \includegraphics[width=0.33\textwidth]{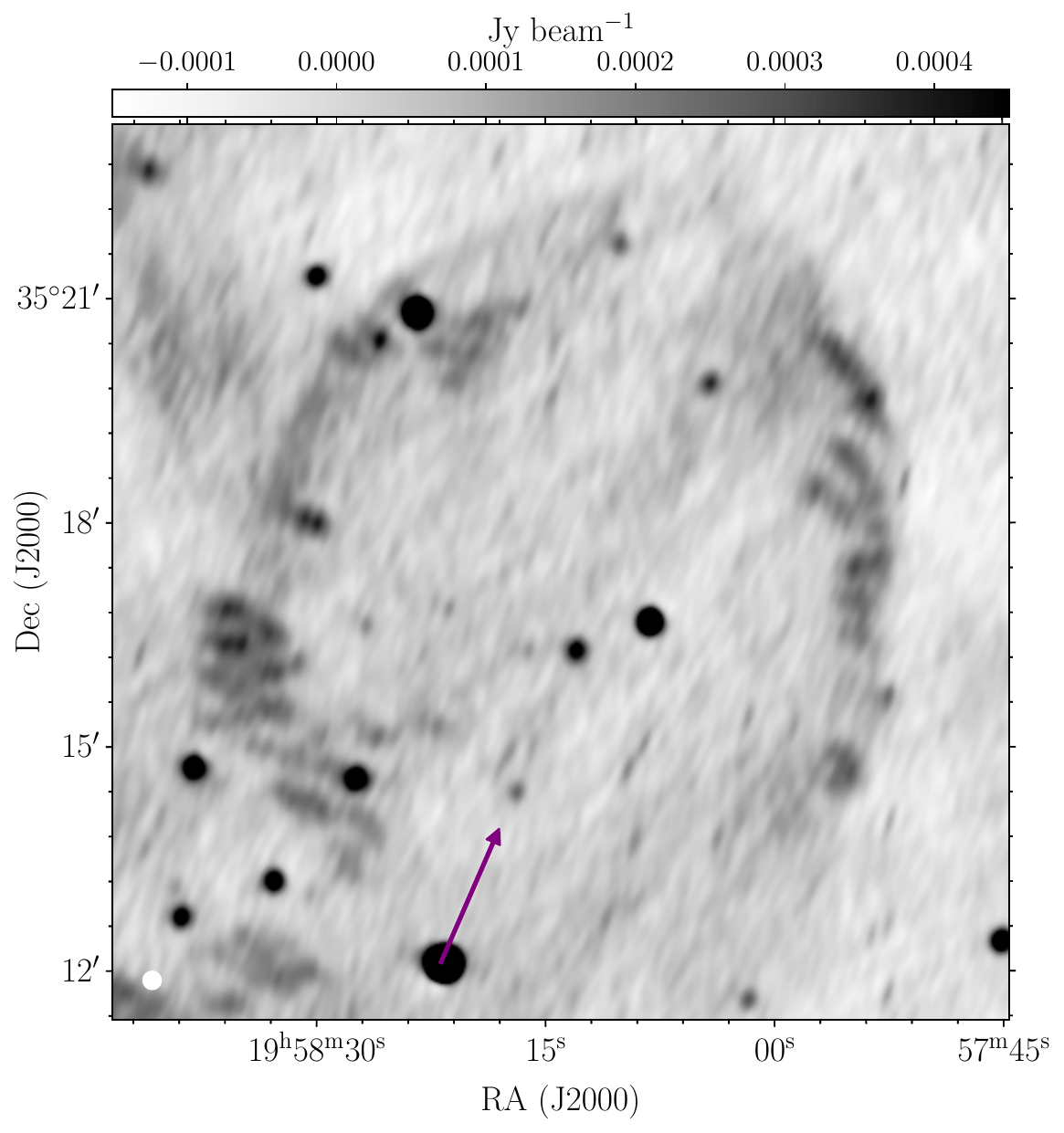}
    \includegraphics[width=0.35\textwidth]{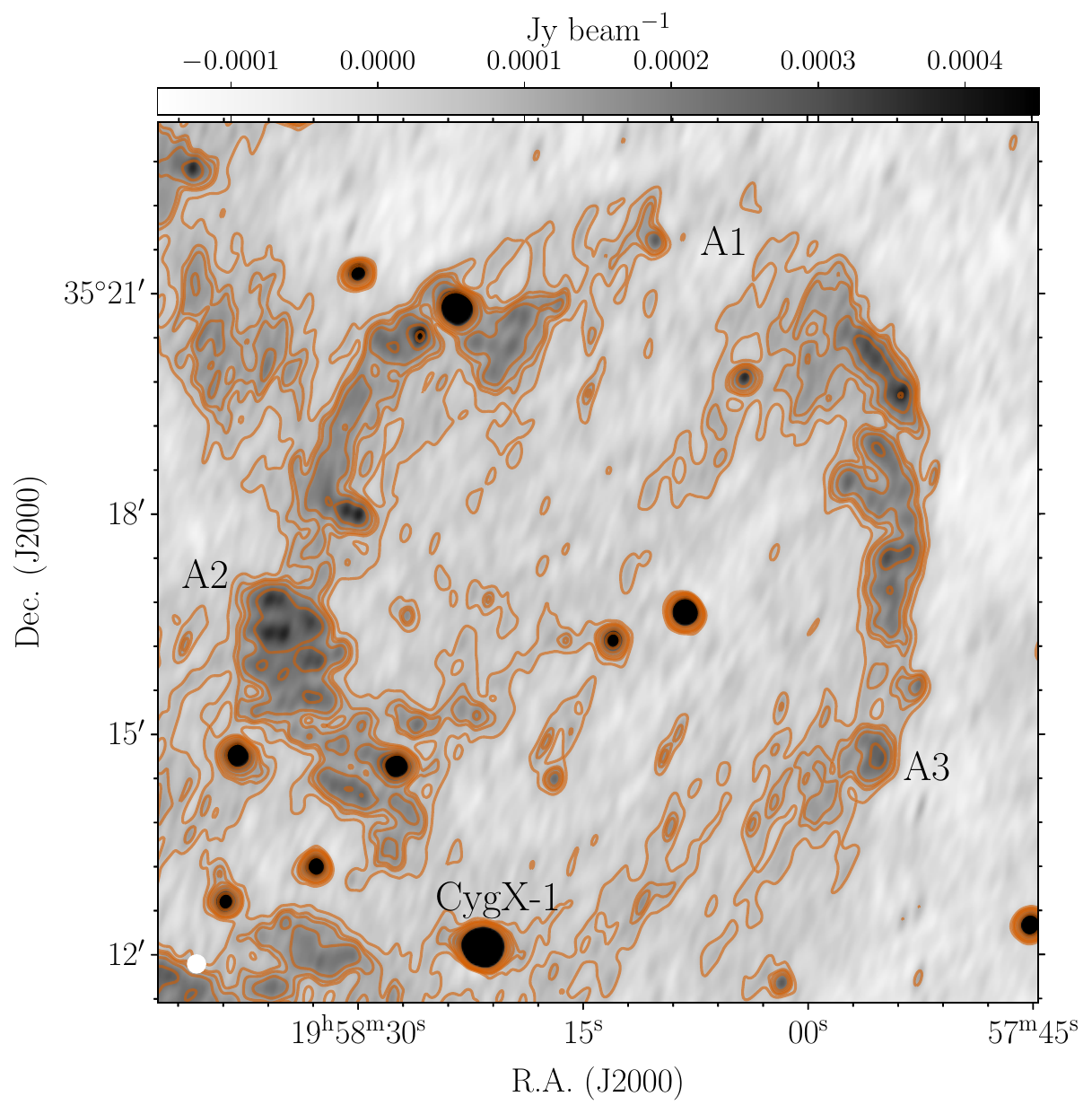}
    \includegraphics[width=0.35\textwidth]{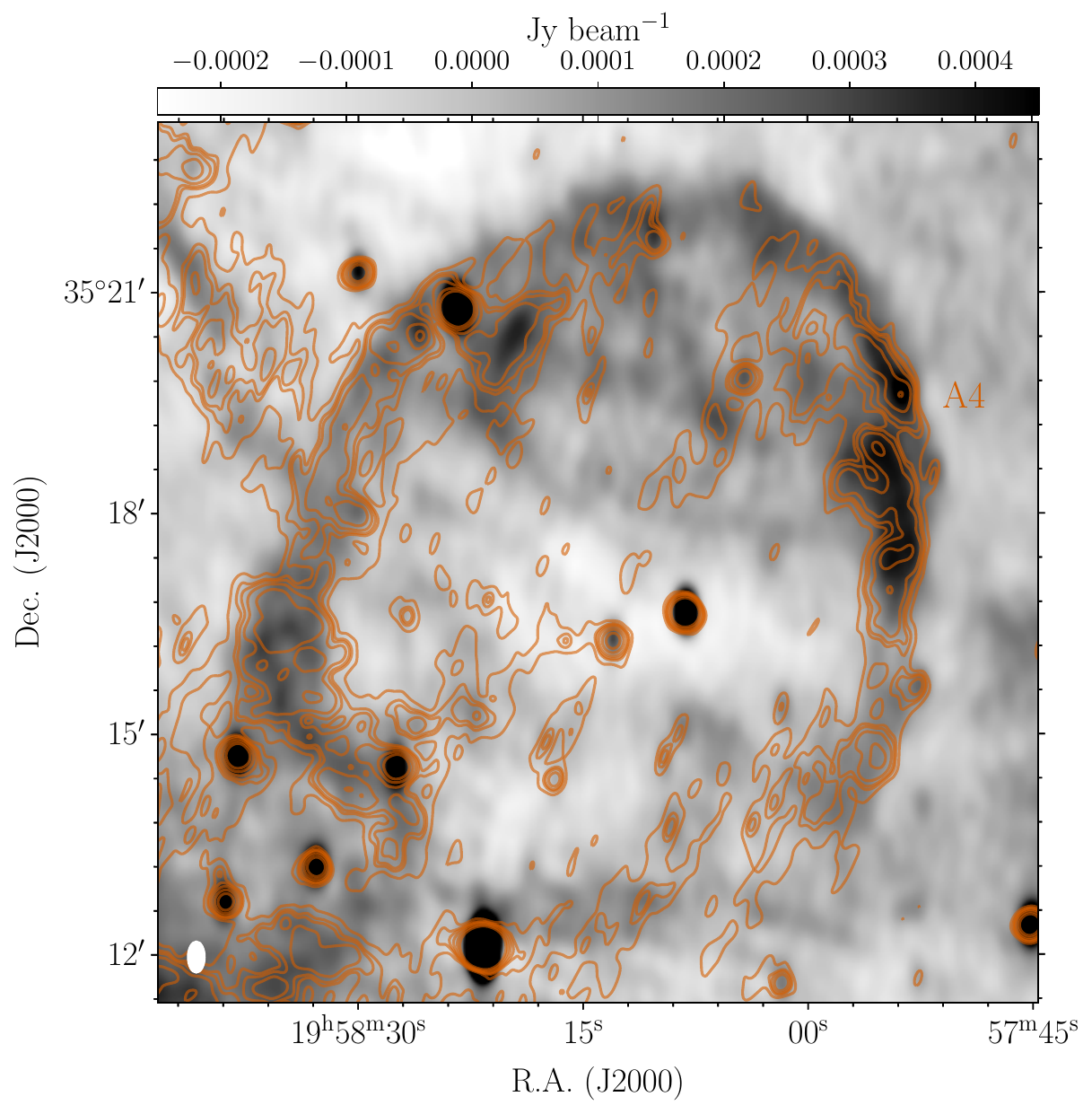}\\
    \end{tabular}
    \caption{Zoomed-in image of the Cyg\,X-1 bow shock as observed by MeerKAT (left and middle panel) and by WSRT (G05; right panel). Left panel: Image of Cyg\,X-1 and the bow shock. The\ purple arrow represents the direction of the hard state jet of Cyg\,X-1 as resolved by \citet{Stirling2001}. Middle panel: L-band MeerKAT image of the bow shock overlaid with orange contours to show the significant emission region in the bow shock. The contours start at a 2$\sigma$ level and increase in steps of $\sigma$, where $\sigma$ is the rms. Right panel: Bow shock contours mapped using MeerKAT L-band data overlaid onto the WSRT radio map of the bow shock in G05 to compare the bow shock over a time span of $\approx$15\,years. }
    \label{fig:zooms_Lband}
\end{figure*}

In Fig. \ref{fig:full_map_Lband} we present the final L-band image obtained by following our data reduction procedure as laid out in Sect. 3. Cyg\,X-1 is seen as a compact, bright source with a flux density of $\approx$ 13\,mJy near the centre of the field and a clear U-shaped arc is identified to the north of Cyg\,X-1 (Fig. \ref{fig:zooms_Lband}, left panel), which is similar to the bow shock identified near Cyg\,X-1 by G05. There is diffuse emission detected to the south of Cyg\,X-1 too (W2 in Fig. \ref{fig:full_map_Lband}; see discussion of `The Whale'), although it is unlikely that this is associated with Cyg\,X-1 given the scale of the emission compared to the size of the northern bow shock. \par

To estimate the parameters of the detected bow shock, we made a contour map of the bow shock region as shown in Fig. \ref{fig:zooms_Lband}, middle panel. The orange contours in the image start at the 2$\sigma$ level and increment in steps of $\sigma$, where $\sigma$ is the rms noise in the region around the bow shock. 
The emission region marked by these contours will be referred to as bow shock region R1 in the rest of the analysis in this paper. The average brightness measured in R1 is 112$\pm$24\,$\mathrm{\mu}$Jy\,bm$^{-1}$, and the integrated flux density of the region (after subtracting compact background emission sources) is 50$\pm$2\,mJy. The leading edge of the bow shock region is marked as A1 in Fig. \ref{fig:zooms_Lband} (middle panel) and is located at an angular separation of 10.8\,arcmin ($L_{l}$ in the formulations in the following sections) from Cyg\,X-1, which corresponds to 6.8\,pc for the distance of Cyg\,X-1. The angular separation between the edges of the bow shock region closest to Cyg\,X-1 (regions marked as A2 and A3) is assumed as the diameter of the radio lobe and is $R_{l}=$\,7.3\,arcmin (4.6\,pc). R1 has varying thickness in different parts of the shock region, so we estimated the average thickness by measuring the width at 20 different areas of the bow shock. The average thickness of the bow shock region ($\Delta R_{l}$) was estimated to be 1.5$\pm$0.2\,arcmin (0.9\,pc). We also overplot R1 contours on the image of the bow shock from G05 to get a comparison of the scales and possible changes in the morphology of the structure (Fig. \ref{fig:zooms_Lband}, right panel). This comparison will be discussed in Sect. \ref{Section 5}.
\subsection{S band}
\begin{figure*}
\centering
\includegraphics[width=0.8\textwidth]{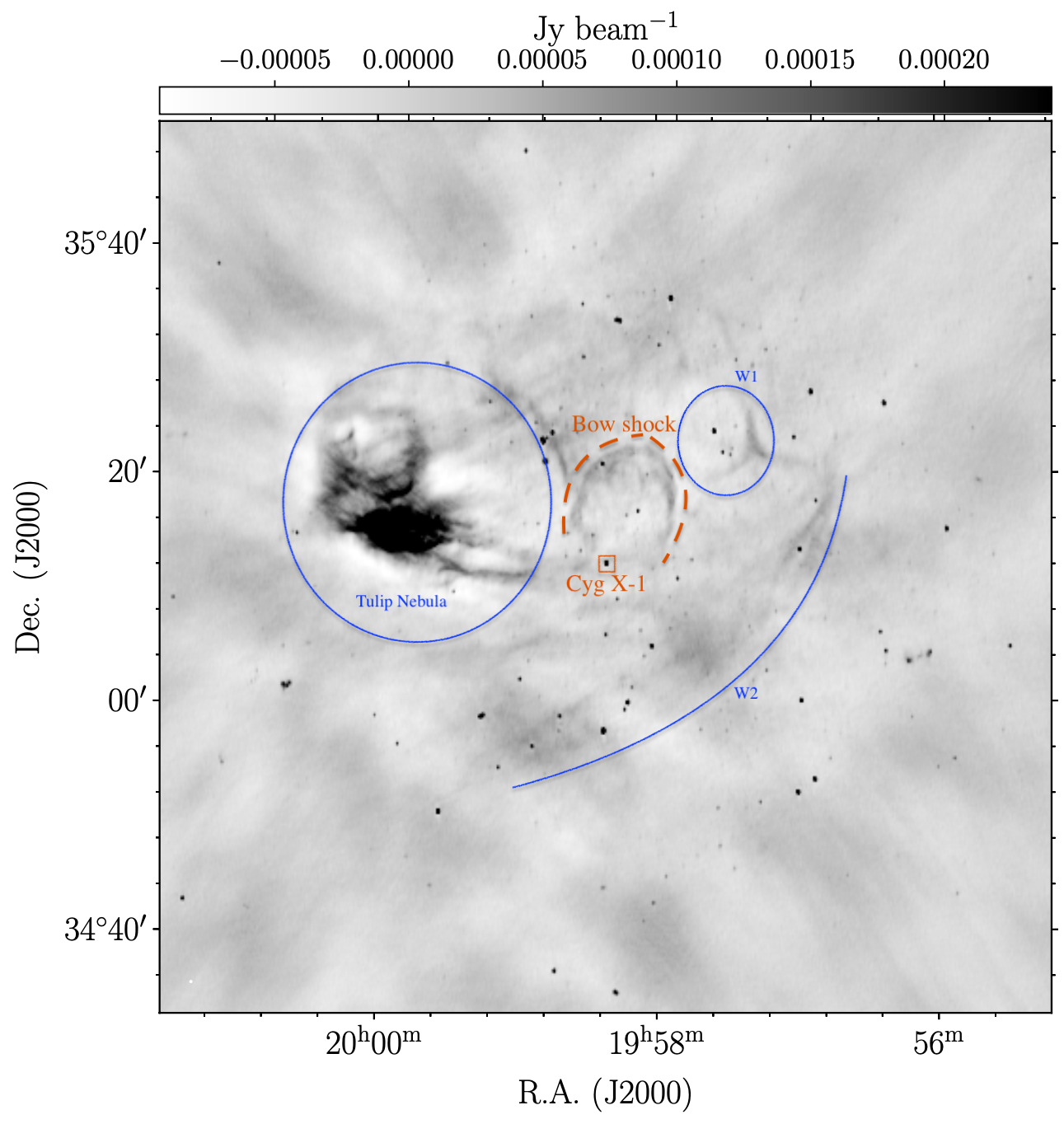}
\caption{Cyg\,X-1 field as observed by the MeerKAT telescope in the S band. Like Fig. \ref{fig:full_map_Lband}, the location of Cyg X-1 is marked with an orange square and the bow shock structure is marked with a dashed orange arc. The Tulip Nebula, W1, and W2 are all also detected in the S band.}
\label{fig:full_map_S_band} 
\end{figure*}

\begin{figure*}
    \centering
    \begin{tabular}{ccc}
    \includegraphics[width=0.33\textwidth]{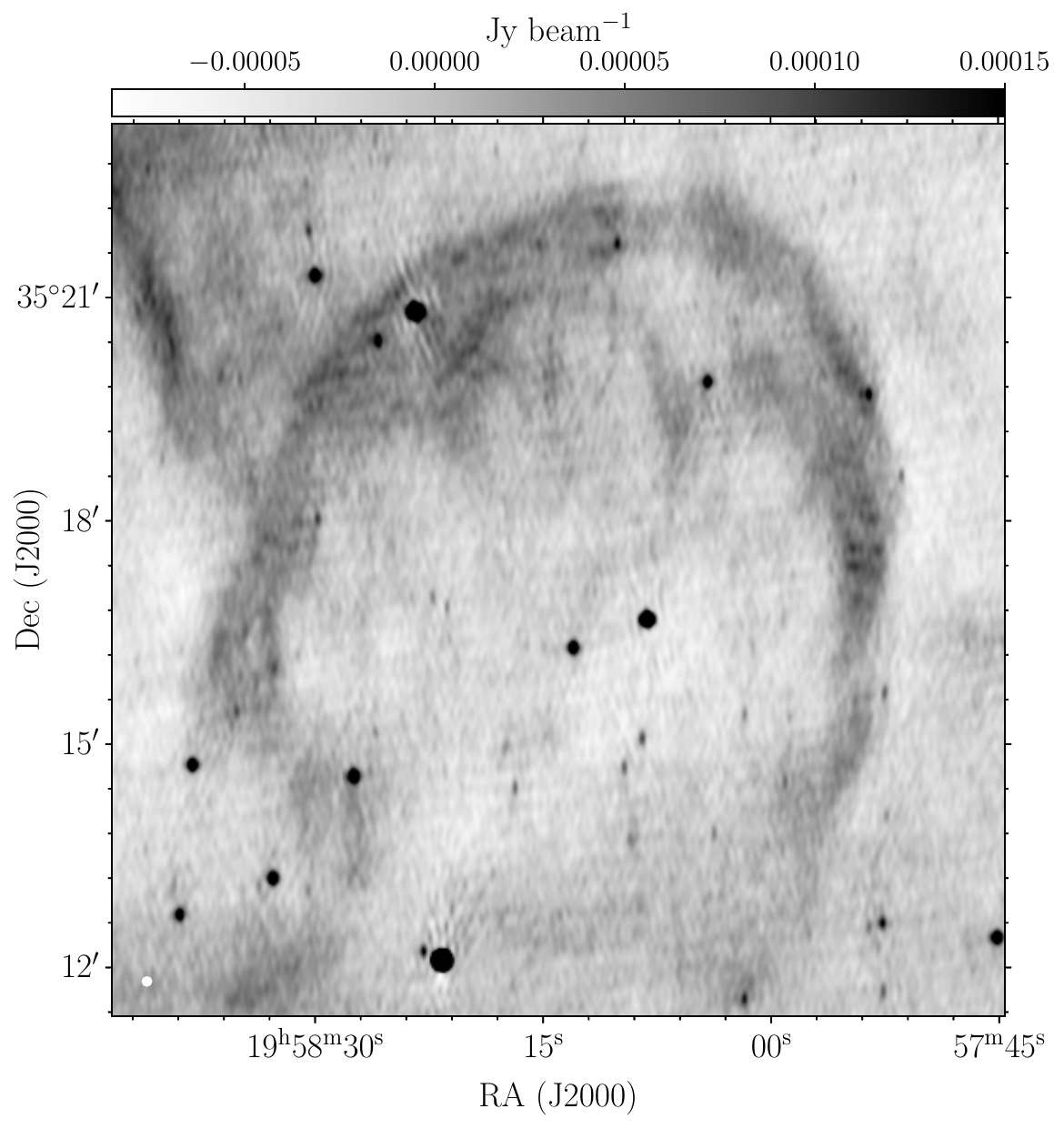}
    \includegraphics[width=0.35\textwidth]{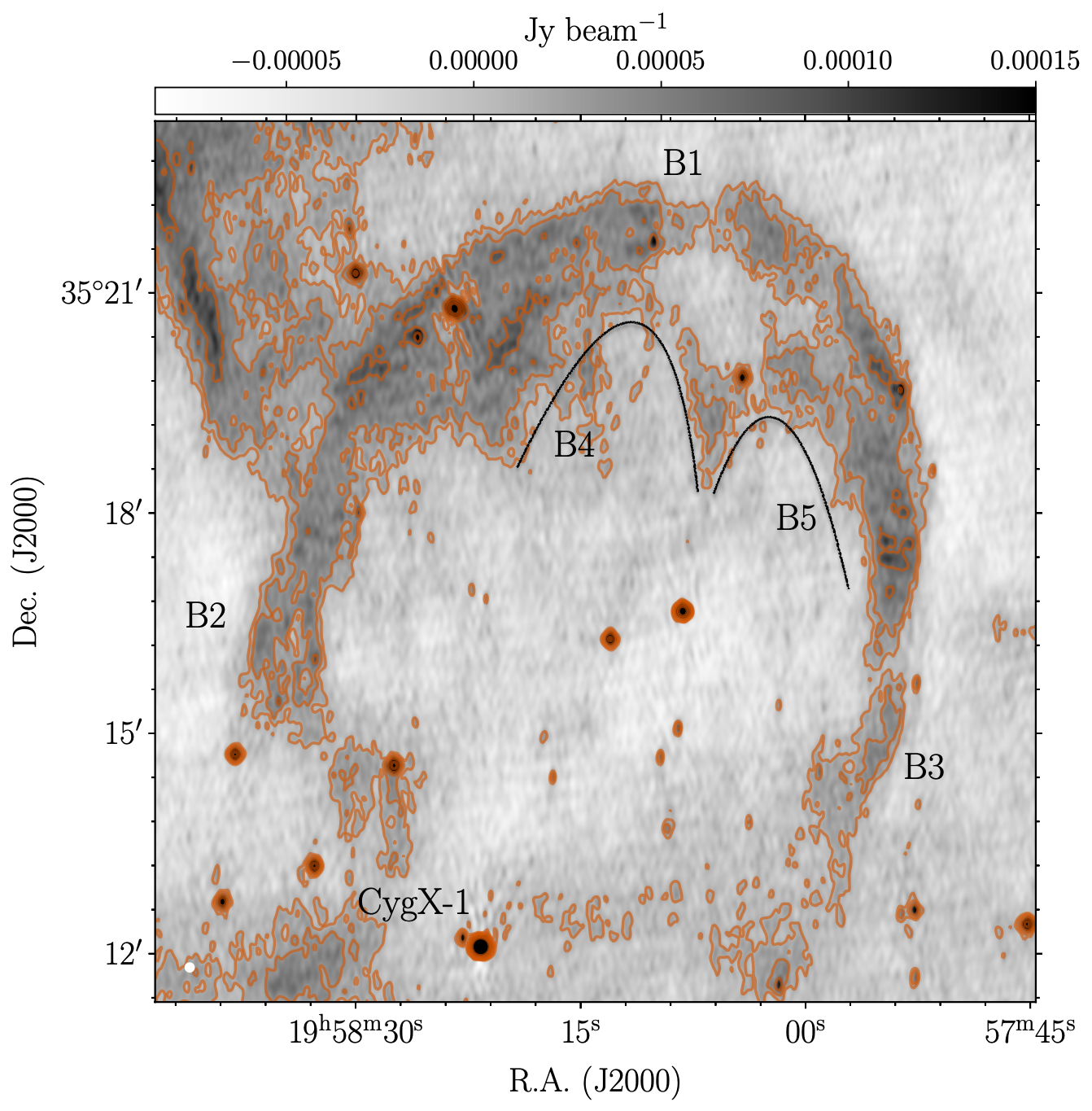}
    \includegraphics[width=0.35\textwidth]{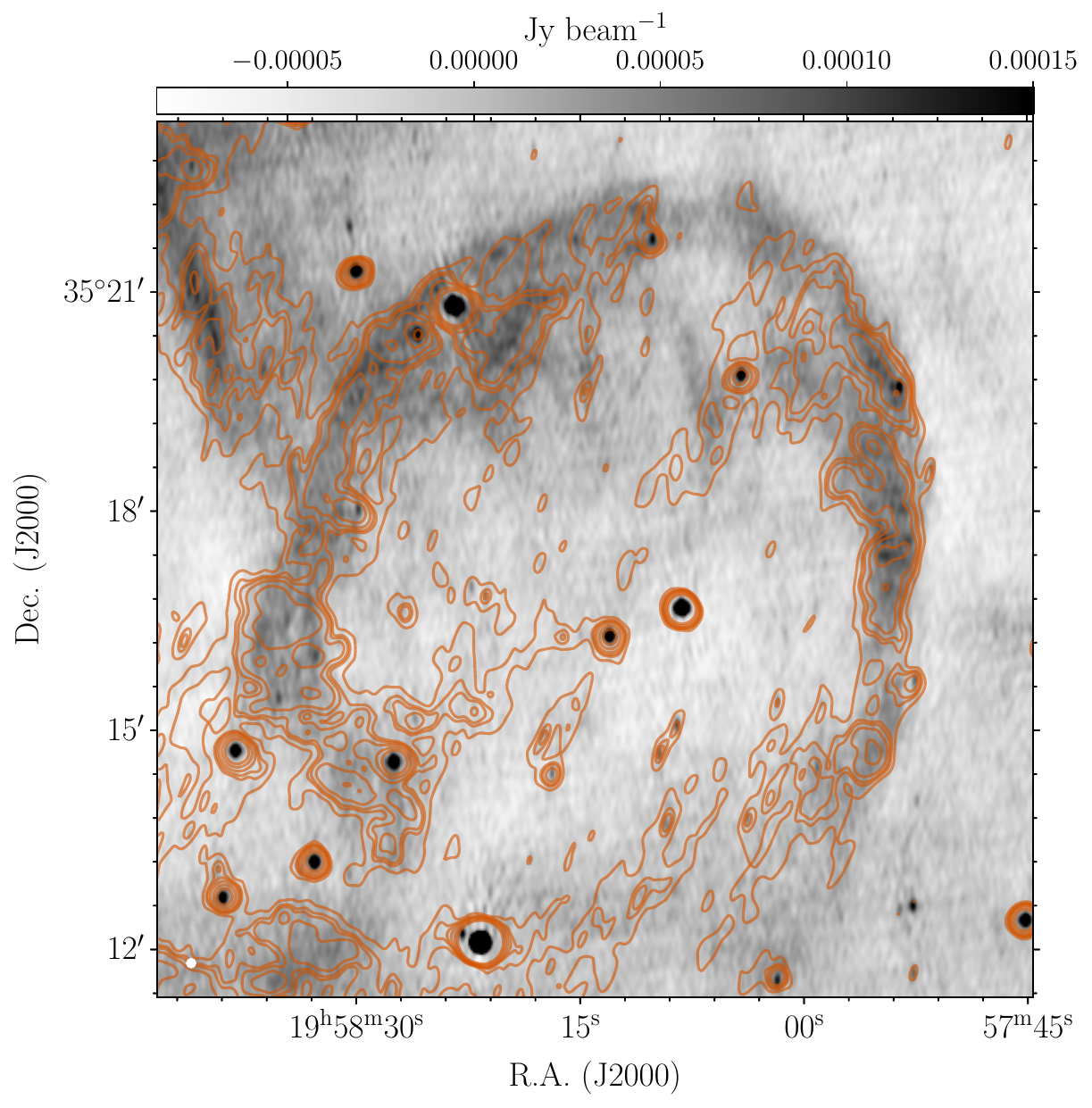}\\
    \end{tabular}
    \caption{Zoomed-in image of the Cyg\,X-1 bow shock as observed by MeerKAT in the S band. Left panel: Image of Cyg\,X-1 and the bow shock. Middle panel: Contours of the S-band emission overlaid on the image, with B1, B2, and B3 marked to show the emission regions used to estimate various bow shock parameters. B4 and B5 mark the location of the smaller arcs within the larger bow shock structure. We also trace these structures with black arcs to draw the reader's eye to the structure. Right panel: L-band bow shock contours overlaid on the S-band image of the bow shock to compare the morphology of the bow shock in the two frequency bands. }
    \label{fig:zooms_Sband}
\end{figure*}

Figure \ref{fig:full_map_S_band} shows the image of our S-band observations' full field of view. Cyg\,X-1 is detected as a compact source with a flux density of 14\,mJy. The bow shock is detected to the north of Cyg\,X-1 and a zoom-in into the bow shock is plotted in Fig. \ref{fig:zooms_Sband}. W1 and W2 are also detected in the S band; however, W2 is contaminated by side lobes, making it challenging to calculate the brightness of this region. These side lobes can be suppressed by making a $uv$ cut at 500 wavelengths, pointing towards short baseline error that is uncalibrated. The resulting image is reported in Fig. \ref{fig:full_map_S_band_uv_cut}. Since the side lobes do not interfere with the location of the bow shock, we derived the bow shock parameters in the S band before applying the $uv$ cut to maintain optimal sensitivity and resolution. We made a contour map focused on the Cyg\,X-1 bow shock as seen in Fig. \ref{fig:zooms_Sband}, middle panel. The orange contours start at 2$\sigma$, and increase by $\sigma$ to a value of 1.27$\times$10$^{-3}$\,Jy\,beam$^{-1}$, where $\sigma$ is 10\uJy and is the rms. 
The average brightness of the bow shock is estimated by considering the emission within the faintest contour level around the bow shock as the edge of the bow shock and is 38$\pm$10\,\uJy. The integrated flux density of the bow shock is 51$\pm$1\,mJy. The angular separation between the leading edge of the bow shock (B1 in Fig. \ref{fig:zooms_Sband}, middle panel) and Cyg\,X-1 is 10.5\,arcmin ($L_{s}$ in the formulations in the following sections). This separation translates to 6.7\,pc for the distance of Cyg\,X-1 and agrees with $L_{l}$. The edges of the bow shock closer to Cyg\,X-1 are marked as B2 and B3 in Fig. \ref{fig:zooms_Sband}, middle panel. The angular separation between these regions is the diameter of the bow shock ($R_{s}$) and is measured to be 7\,arcmin (4.4\,pc). The width of the bow shock ($\Delta R_{s}$) is calculated in the same way as the L-band observations, where we averaged the width of the bow shock in 20 areas of the bow shock and the mean width is 1.4$\pm$0.2\,arcmin, which is consistent with $\Delta R_{l}$.

\begin{table*}
\caption{Known, measured, and estimated parameters of the Cyg\,X-1 bow shock.}            
\label{tab:BHpar}     
\centering                      
\begin{tabular}{c c c}        
\hline                     
\multicolumn{3}{c}{Known parameters of Cyg\,X-1 from literature}  \\  
\hline
Parameter & Value & Reference \\  
\hline
Distance & 2.22$^{+0.18}_{-0.17}$\,kpc & [1] \\
Jet inclination angle & 27.18$\pm$0.18$^\circ$ & [1] \\
De-reddened H$_\alpha$ flux of the bow shock (S$_{\rm H\alpha}$) & 1.3$\times$10$^{-16}$\,ergs cm$^{-2}$\,arcsec$^{-2}$ & [2, 3] \\
Jet opening angle ($^{a}$) & 0.2$^\circ$ -- 2$^\circ$ & [4] \\
Width of the H$_\alpha$ ring ($\Delta$) & 1.6$\times$10$^{18}$\,cm & [2] \\
\hline                                       
\multicolumn{3}{c}{Bow shock parameters using L-band map}  \\ 
\hline
\multicolumn{1}{c}{} & Measured values & Physical values \\  
\hline
Separation between Cyg\,X-1 and apex of bow shock ($L_{l}$) & 10.8\,arcmin & 6.8\,pc \\
Separation between closest edges of bow shock ($R_{l}$) & 7.3\,arcmin & 4.6\,pc \\
bow shock width ($\Delta R_{l}$) & 1.5$\pm$0.2\,arcmin & 0.9\,pc \\
Projected area of the bow shock & 97771\,arcsec$^{2}$ & 11\,pc$^{2}$ \\
Volume of the bow shock & --- & 10\,pc$^{3}$ \\
Mean bow shock brightness & 110$\pm$24\uJy & --- \\
\hline                      
\multicolumn{3}{c}{Bow shock parameters using S-band map}  \\ 
\hline
\multicolumn{1}{c}{} & Measured values & Physical values \\  
\hline
Separation between Cyg\,X-1 and apex of bow shock ($L_{s}$) & 10.5\,arcmin & 6.7\,pc \\
Separation between closest edges of bow shock ($R_{s}$) & 7.0\,arcmin & 4.4\,pc \\
Bow shock width ($\Delta R_{s}$) & 1.4$\pm$0.2\,arcmin & 0.9\,pc \\
Projected area of the bow shock & 95689\,arcsec$^{2}$ & 11\,pc$^{2}$ \\
Volume of the bow shock & --- & 9\,pc$^{3}$ \\
Mean bow shock brightness & 38$\pm$13\uJy & --- \\
\hline                      
\multicolumn{3}{c}{Properties of the bow shock and the ISM}  \\ 
\hline   
Electron density (L band) (n$_\mathrm{eL}(T)$) & \multicolumn{2}{c}{$24^{+12}_{-8}$ cm$^{-3}$} \\
Electron density (S band) (n$_\mathrm{eS}(T)$) & \multicolumn{2}{c}{$28^{+14}_{-9}$ cm$^{-3}$} \\
Electron density (H$_\alpha$ maps) (n$_{eh}$) & \multicolumn{2}{c}{$11 \pm 2$ cm$^{-3}$} \\
Velocity of leading edge of bow shock ($\dot{L}$) & \multicolumn{2}{c}{21\,km\,s$^{-1}<\dot{L} <\,364$\,km\,s$^{-1}$} \\
Age of the jet & \multicolumn{2}{c}{0.04--0.3\,Myr} \\
Jet power (Q$_{jet}$$^{b}$; Jet opening angle 0$^\circ$ -- 30$^\circ$ ) & \multicolumn{2}{c}{2$\times$10$^{34}$\,ergs s${^{-1}}<$ Q$_{jet}^{b}<$ 1$\times$10$^{37}$\,ergs s$^{-1}$} \\
Jet power (Q$_{jet}$$^{a}$) & \multicolumn{2}{c}{ 2$\times$${10^{31}}$\,ergs s$^{-1}<$ Q$_{jet}^{a}<$ 1$\times$10$^{35}$\,ergs s$^{-1}$} \\
Pressure inside the bow shock & \multicolumn{2}{c}{3$\times$10$^{-13}$ -- 2$\times$10$^{-11}$\,ergs\,cm$^{-3}$} \\
\hline                             
\end{tabular}
\tablefoot{The superscript a and b represent the estimates using two different opening angle ranges.}
\tablebib{[1] \citet{Miller-Jones2021}, [2] \citet{Russell2007}, [3] \citet{Sell2015} and [4] \citet{Stirling2001}.}
\end{table*}

\section{Discussion}\label{Section 5}

\subsection{Spectral index}

\begin{figure*}
    \centering
    \begin{tabular}{cc}
    \includegraphics[width=0.45\textwidth]{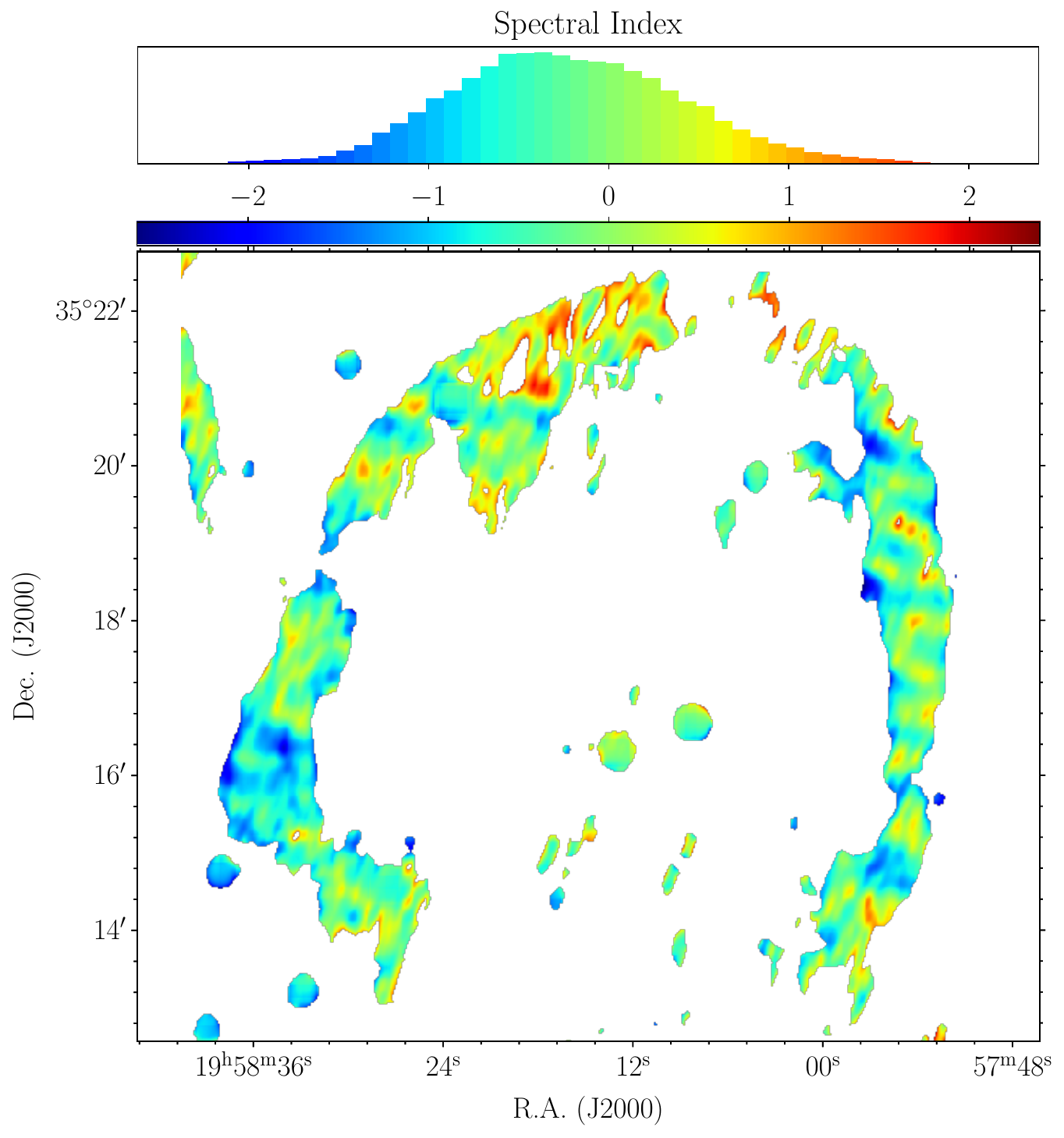}
    \includegraphics[width=0.45\textwidth]{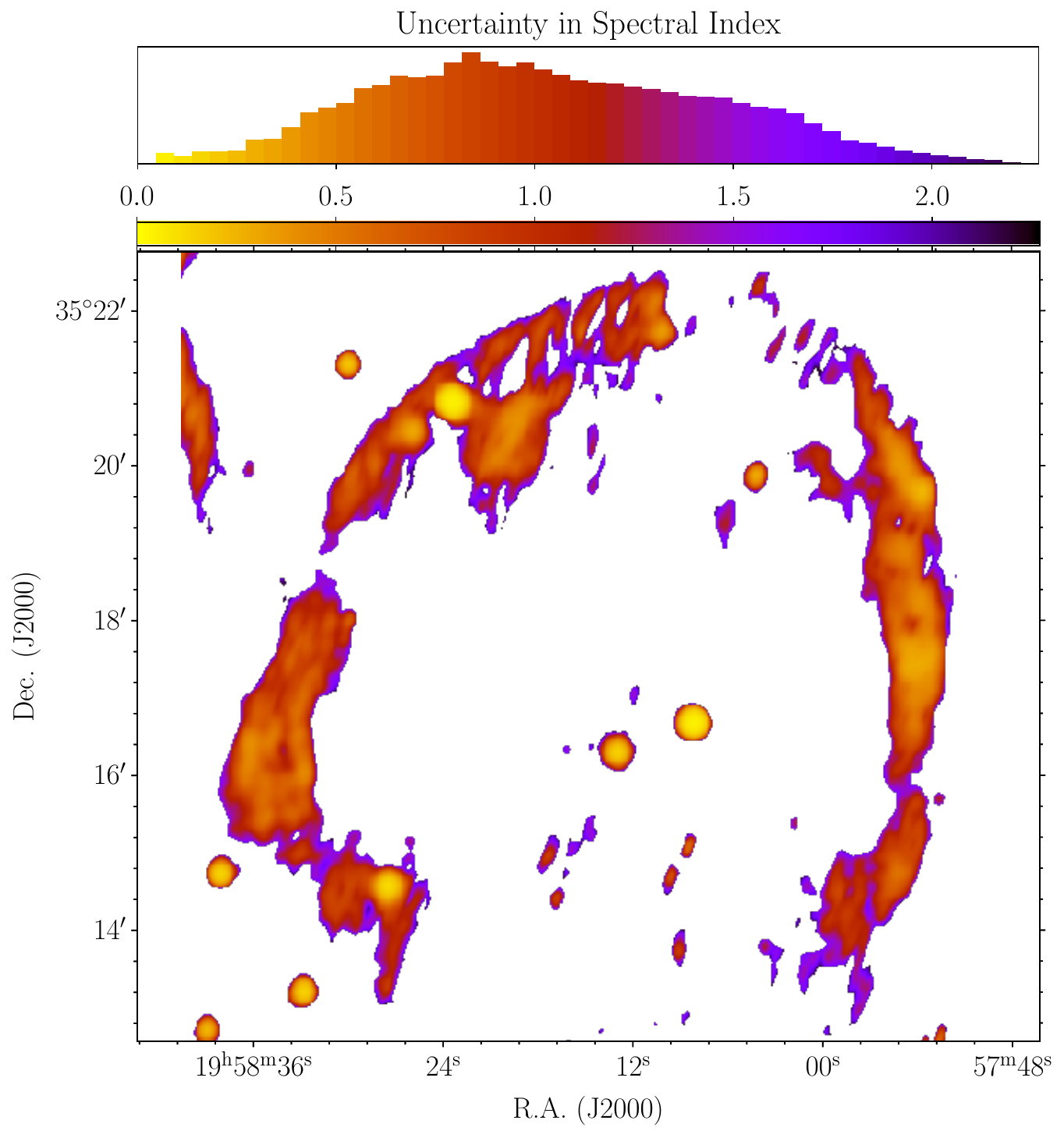}
    \end{tabular}
    \caption{Spectral index and error distribution map of the bow shock structure between L- and S-band MeerKAT detections. Bottom panels: Distribution in the RA and Decl plane. Top panels: Histogram representations of the distribution of value and errors of the spectral index. The spectral index does not vary noticeably across the bow shock, and the values lie between -0.9 and 0.4, with the error bars consistent with 1. }
\label{fig:spectral_map} 
\end{figure*}

Particles accelerated due to interaction with the ISM have a power-law energy distribution at the shock front that is seen as a power-law radio spectrum ($S_{\nu}\,\propto\,\nu^{\alpha}$, where $\alpha$ is the spectral index). The electron population in these regions can emit non-thermally, primarily via the synchrotron process or thermally via Bremsstrahlung processes. The electron density, the particle acceleration process and its efficiency, and the temperature of the shock determine the emitting process of the shock, and in turn the spectral index. A flat spectral index indicates thermal emission processes whereas a steep spectral index suggests that synchrotron emission processes contribute to the bow shock emission \citep{Longair2011}. Here, we investigate the integrated spectral index and the spectral index distribution of the bow shock to understand the origin of the emission.  
We compared the measured flux densities of the bow shock detected in the two frequency bands of the MeerKAT data. The $uv$ plane in the L band has a denser coverage of the shorter spacings as compared to the S band, and thus we forced the same $uv$-minimum cut in both the L- and S-band image to recover similar angular scales at the two frequencies. We then convolved the S-band image to the resolution of the L-band image (14\,arcsec) so we can compare the images with the same resolution and minimise resolving out flux in the S band. We also only map the spectral index of emission that was greater than 2$\sigma$, where $\sigma$ is 15\,\uJy  and 20\,\uJy for the L-band and low-resolution S-band images, respectively. \par
In Fig. 5, we present the spectral index distribution of the Cyg\,X-1 bow shock (left panel) and the right panel represents the error in the spectral index measurement. We have also plotted the histogram of the spectral index and spectral index error distribution. The spectral index is mostly smooth across the bow shock and the 68$\%$ confidence interval of the values is -0.9$\leq \alpha \leq$ 0.4. The spectral index error distribution has a 68$\%$ confidence interval of 0.6$\leq \Delta\alpha \leq$ 1.5. Thus, the error bars on the spectral index map are too large to conclude whether the emission in different regions of the bow shock is flat or negative spectrum as it appears to be consistent with both. Even though the absolute spectral index values have large uncertainties, we can see that there are no noticeable spectral index gradients, pointing to the same emission mechanism for the complete structure. The integrated flux density of the bow shock is measured as 35$\pm$3\,mJy and 27$\pm$3\,mJy, leading to an integrated spectral index of -0.4$\pm$0.5. It is worth noting that these integrated flux densities are significantly different than the integrated flux density measured before applying the $uv$-cut and forcing the same beam size in both bands. We find that the integrated spectral index is also inconclusive to determine the origin of the bow shock emission. To determine the origin of the emission from the radio spectral index alone, we need a more accurate spectral index measurement, which will be possible in the future with our Ultra High-Frequency band observation of this region that covers a frequency range 544--1087\,MHz using the MeerKAT. 

\subsection{Modelling the jet--ISM interaction site}\label{temp-density}

The bright jet--ISM interaction site as seen in the MeerKAT L-band and S-band images provides an opportunity to estimate the time-averaged power being pumped into the region by the jets of Cyg\,X-1. Due to the lack of models that directly explain BHXB jet and ISM interactions, we followed a relatively simplistic model of extending the AGN jet--IGM interaction expectations to the BHXB case. This follows the model explored by \citet{Kaiser1997} and \citet{Kaiser2004} for AGNs and BHXBs, respectively. They use the surface brightness of the jet--ISM interaction site to determine various parameters of the jet and the radio bubbles, extending it to the specific case of a BHXB, GRS\,1915+105. A similar, though simplified, model was applied to the bow shock detected around Cyg\,X-1 by \citet{Gallo2005}. In this paper, we extend that model and use it to explain the new L-band and S-band detections of the bow shock. There are a few notable differences between the modelling input parameters in G05 and our work, which have been described in the following section. Here, we present detailed calculations of the ISM density of the surroundings, the expansion velocity of the bow shock, the age of the jet flow and the energy transport rate that has been able to create the bow shock. We have also provided the Jupyter Notebook and thus the Python code that we used to do these calculations (see Sect. \ref{Section 8}). 
We begin by summarising the input parameters and assumptions of the model. We explain the choices when these parameters are mentioned in the relevant sections:

\begin{itemize}
    \item The timescale of the variability of the BHXB is small as compared to the dynamical change timescale of the large-scale radio lobes.
    \item The bow shock radio emission is of Bremsstrahlung origin and the assumed temperature range of the emitting region is $10^4<T<3\times10^6$K (G05). 
    \item The ISM gas producing the radio emission is completely ionised (ionisation fraction is 1).
    \item The bow shock is as deep in our line-of-sight as it is wide in the plane of the sky ($\Delta R$).
    \item The jet opening angle has been assumed to be 0$^\circ$-30$^\circ$ to obtain generalised estimations.
    \item We assumed that the bow shock brightness is the same everywhere within the estimated error bars and thus all estimations are homogeneous across the bow shock.
    
\end{itemize}

The first step in modelling the jet--ISM interaction site is to determine the shocked and unshocked ISM density. For the case of the Cyg\,X-1 bow shock, we used the radio emission detected from the bow shock in the L- and S-band MeerKAT observations and the H$_{\alpha}$ surface brightness of the bow shock \citep{Russell2007,Sell2015}. 

\subsubsubsection{Using radio emissivity}

\begin{figure}
\centering
\includegraphics[width=0.80\textwidth]{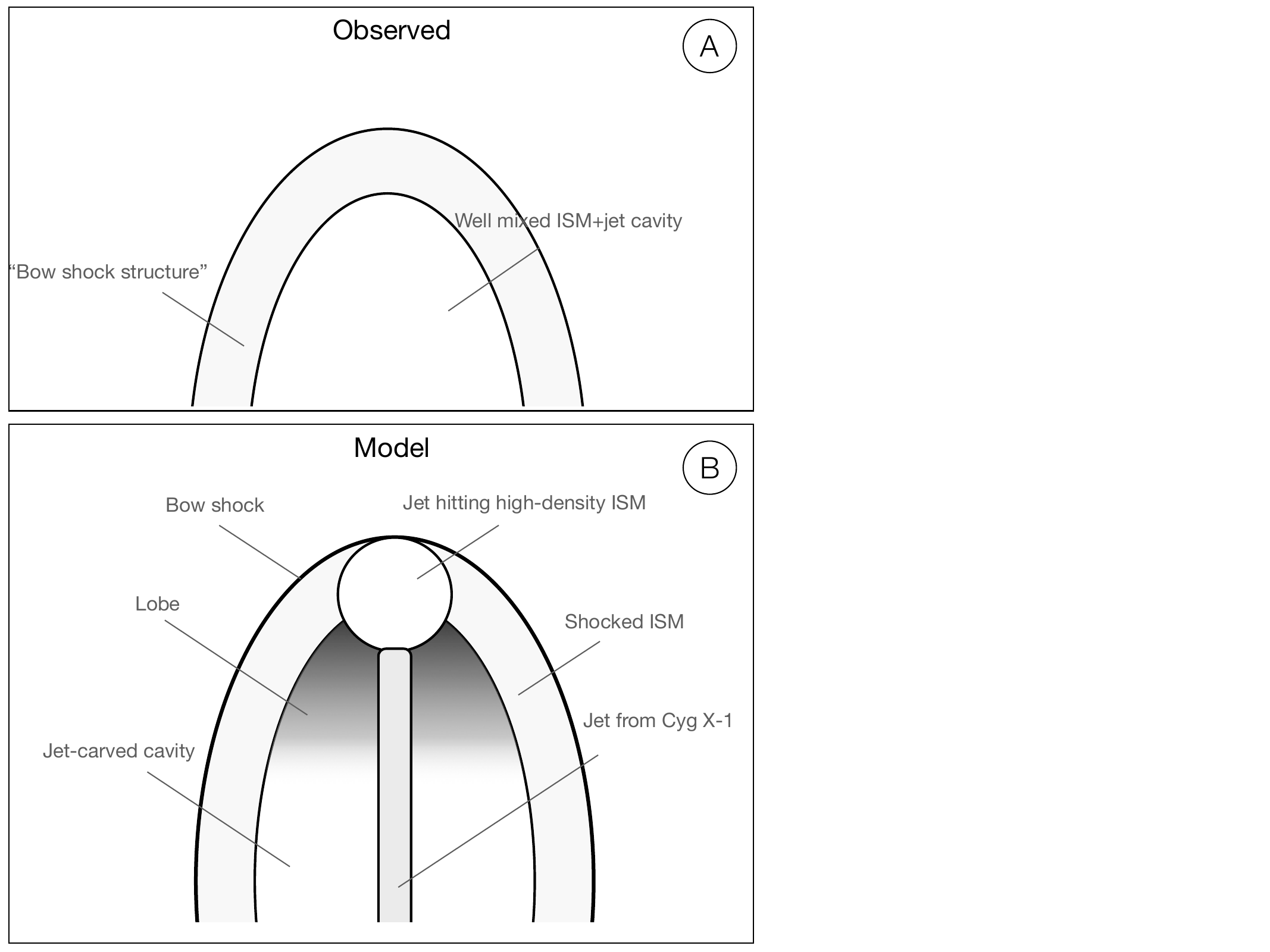}
\caption{\textit{Panel A:} Simplified schematic of the observations of the jet--ISM interaction site near Cyg\,X-1 as detected in the MeerKAT image. \textit{Panel B:} Schematic of the model used to derive the jet parameters of Cyg\,X-1.}
\label{fig:schem}
\end{figure}

\noindent When the radio jets from Cyg\,X-1 plunge into its environment, it creates an over-pressured radio lobe that expands sideways. The over-pressured lobe allows the jet to propagate efficiently with minimal energy losses until it reaches an over-dense medium. As the end of the jet interacts with this over-dense region, it shocks the ISM and transfers energy to the medium. This results in a shock-compressed ISM at the edge of an over-pressured lobe. Due to limb brightening effects, we detect the edges of the shocked fronts in our radio and optical maps. We present a schematic of our model and a simplification that explains what we observe in our radio observations in Fig. \ref{fig:schem}. As presented in the schematic, the only signature of the jet--ISM interaction observed in the case of Cyg\,X-1 is the bow shock structure. We do not detect a distinct end of the jet as it interacts with the ambient medium. We also do not detect the over-pressured lobe. The existence of these signatures is a result of the interplay between the jet power, the environment in which the jet is dissipating energy and the particle content of the plasma. It is not straightforward to disentangle the contribution of these parameters to single out the reason for not detecting hotspots or the over-pressured lobe. The same issue has long been an open question in the extragalactic equivalent of these systems, namely radio galaxies. For instance, the radio jets of Cygnus A are associated with radio bubbles, hotspots and bow shock structure \citep{Carilli1996} whereas no hotspots are detected in the case of Centaurus A \citep{Feain2011}.
For the case of Cyg\,X-1, we speculate that the ISM and the jet-induced cavity are well mixed, such that the jet is not very collimated at the distance where the bow shock structure is detected. This would lead to a more gradual dissipation of kinetic energy than what is needed for creation of a distinct, bright hotspot. This idea is also supported by the detection of a broader, non-uniform bow shock structure instead of a sharp-edged bow shock. 
The analogy with radio galaxies is furthermore seen when comparing Cyg X-1 with the recently detected bow shock driven by the Galactic system GRS 1915+105. GRS\,1915+105 has been associated with a hotspot, an over-pressured bright lobe \citep{Chaty2001,Rodriguez1998}, and a bow shock structure \citep{Motta2025}, making it distinctly different from the behaviour of jets from Cyg\,X-1 at large scales. 

We attempted to measure the integrated radio spectral index and its distribution across the bow shock with the data presented in this paper; however, the measurements are not accurate enough to determine whether the emission can be attributed to Bremsstrahlung processes. Thus, here we relied on the conclusions of \citet{Gallo2005}, who suggested that the surface brightness of the bow shock in their WSRT L-band detection and the H$_{\alpha}$ detection points to the radio emission of the ring-like structure being of Bremsstrahlung origin. The minimum temperature of the bow shock is thus $>10^4$K, below which ionised hydrogen gas is unable to efficiently emit braking radiation. \citet{Gallo2005} could not find any soft X-ray emission from the bow shock region and thus put an upper limit of $3\times10^6$\,K. For our calculations, we considered the temperature range $10^4<$T$<3\times10^6$\,K. \par
The Bremsstrahlung radio emissivity ($\epsilon_\nu$) and monochromatic luminosity ($L_\nu$) of an ionised hydrogen region \citep{Longair1994} is related as
 $$\epsilon_\nu = \frac{L_\nu}{V} =  C_{radio} g(\nu, T) \frac{n_e^2}{\sqrt{T}} \exp{(\frac{h\nu}{k_b T}),} $$
 where 
 
 $ L_\nu = F_\nu \times 4\pi\times d^2 $ is the monochromatic luminosity,

$F_\nu$ is the integrated radio flux density of a flat spectrum source at a central frequency $\nu$,

$d$ is the distance to the flat spectrum source,

$V$ is the volume of the emitting region,

$T$ is the temperature of the shocked ISM,

$ C_{radio} = 6.8\times 10^{-38} \mathrm{erg~s^{-1}~cm^{-3}~Hz^{-1}} $, \\

$
g(\nu, T) \approx \frac{\sqrt{3}}{2 \pi}\left[\ln \left(\frac{128 \epsilon_0^2 k_b^3 T^3}{m_e e^4 \nu^2 Z^2}\right)-\gamma^2\right]$ is the Gaunt factor,

$\epsilon_0 = 1$,

$k_b$ is the Boltzmann constant, $m_e$ is the mass of electron, $e$ is the electric charge, $\nu$ is the central observing frequency, $Z$ is the hydrogen atomic number, and $n_e$ is the electron density of ISM.

 \par
The bow shock structure appears as an arc or a U-shaped curve with thickness $\Delta R$. Knowledge of the true shape of the bow shock is sensitivity limited and suffers from projection effects of the three-dimensional bow shock as a two-dimensional bow shock structure in the image plane. For this paper, we only use the detected radio emission in our observations to estimate the volume and thus emissivity for the bow shock structure. We summarise these values for both the L-band and S-band detection of the bow shock in Table \ref{tab:BHpar}. Rearranging the equation for Bremsstrahlung emissivity, we can derive the electron density as

$$n_e = \sqrt{\frac{L_\nu}{ V g(\nu, T) \sqrt{T} C_{radio}}}.$$
Using the values of the measurements as reported in Table 1, we can estimate the electron density of the shocked ISM for both the L-band and S-band bow shock parameters. Since there are error bars associated with the measurements of the ring thickness, the integrated flux density, distance to Cyg\,X-1, and a range of temperatures that can emit the radiation, we use a Markov Chain Monte Carlo (MCMC) approach to derive a distribution of the electron density of the shocked ISM. The resulting probability density distribution using the bow shock parameters derived from the L-band detection suggests that the electron density of the shocked ISM is $n_\mathrm{eL}(T) = 24^{+12}_{-8}$cm$^{-3}$, representing the median, 16 and 84 percentiles of the distribution. Similarly, using the bow shock parameters from the S-band detection we estimate the electron density of the shocked ISM to be $n_\mathrm{eS}(T) = 28^{+14}_{-9}$cm$^{-3}$. Such high shocked ISM densities have been observed in regions around two other XRBs; Vela\,X-1 \citep{Eijnden2022} and the newly reported GRS\,1915+105 \citep{Motta2025}. \par
At high temperatures of $>10^4$\,K, almost all the hydrogen will be ionised at such densities and thus the ionisation fraction will be unity. The total number density of the shocked ISM will then be the same as the electron density, $n_\mathrm{eL}(T)$ and $n_\mathrm{eS}(T)$. This is a notably different assumption as compared to the one used in G05, resulting in a different total number density of the shocked ISM as compared to the one reported in G05. For strong shocks and assuming monatomic gas, shocked ISM is four times denser than the pre-shocked ISM \citep{Landau1991}, the pre-shocked ISM number density will be $\approx \,6\,$cm$^{-3}$ when considering the L-band bow shock parameters and $\approx \,7\,$cm$^{-3}$ when considering S-band bow shock parameters. 
\subsubsubsection{Using the H$_{\alpha}$ surface brightness}

\noindent The Cyg\,X-1 bow shock has been the subject of radio and optical campaigns to understand the energetics responsible for creating the bow shock structure. \citet{Russell2007} reported the detection of the bow shock structure using an H$_{\alpha}$ filter, which supported a thermal plasma origin for the emission seen in the radio and optical bands. Similar to radio surface brightness, the H$_{\alpha}$ surface brightness can also be used to determine the electron density of the emitting region. We used the following relation between H$_{\alpha}$ surface emissivity ($j_{\rm H\alpha}$), the temperature of the emitting region (T$_{h}$), and the electron density ($n_{\mathrm{eh}}$) to estimate the possible range of the electron density of the shocked gas that can explain the detected H$_{\alpha}$ emission \citep{Osterbrock1989, MacKey2013, Gvaramadze2018, Eijnden2022}: 
\begin{equation}
    j_{\rm H\alpha} = 2.85\times10^{-33}\text{erg s}^{-1}\text{ cm}^{-3}\text{ arcsec}^{-2} \left( \frac{T}{\text{K}} \right)^{-0.9}\left(\frac{n_{\mathrm{eh}}}{\text{cm}^{-3}}\right)^2 \label{eq1}
.\end{equation}The surface brightness ($S_{\rm H\alpha}$) that is measured from H$_{\alpha}$ maps is a result of integrating the surface emissivity along the line of sight in the bow shock and is given by

$$S_{\rm H\alpha} = \int j_{\rm H\alpha}(x) dx = j_{\rm H\alpha} \Delta.$$
Rearranging these equations, we can use the surface brightness and temperature range to get an ISM density distribution:

$$n_{\mathrm{eh}}(T) = \sqrt{S_{\rm H\alpha}\frac{(T/\text{K})^{0.9}}{C_{\rm H\alpha}\Delta}}, $$ 
where $C_{\rm H\alpha} = 2.85\times10^{-33}\text{erg s}^{-1}\text{ cm}^{-3}\text{ arcsec}^{-2}$ and $\Delta$ is the width of the bow shock in cm. The background subtracted H$_\alpha$  surface brightness of the Cyg\,X-1 bow shock is $\approx$ 1.3$\times$10$^{-16}$\ergs cm$^{-2}$\,arcsec$^{-2}$ \citep{Sell2015}. We limit temperature to a conservative range of 5$\times$10$^3$--2$\times$10$^4$\,K motivated by the range of temperatures for which Eq. \ref{eq1} is valid. We note that the temperature range is significantly lower than the consideration for the Bremsstrahlung emitting region to ensure that all hydrogen is not ionised and can still emit H$_\alpha$. Using the above temperature range and a 15$\%$ error budget for S$_{\rm H\alpha}$ and $\Delta$, we perform MCMC simulations to obtain a distribution of $n_{\mathrm{eh}}$. We find that the ISM densities that can explain the detected H$_\alpha$ emission can be constrained to $11 \pm 2\,cm^{-3}$, where the error bars are the 16th and 84th percentile of the distribution.

\subsubsubsection{Combining radio and $\mathrm{H\alpha}$ maps}
The shocked ISM density determined above using the radio detections in the L and S bands is 2-3 times higher than the ISM density that can explain the emission detected in the H$_\alpha$ band. The possible temperature ranges combined with different ISM densities suggest that the bow shock structure could be a result of multiple shocks, with a contact discontinuity between the radio and H$_\alpha$ emitting region to be a factor of 2-3 ($n_{\mathrm{eL}}$/$n_{\mathrm{eh}}$ to $n_{\mathrm{eS}}$/$n_{\mathrm{eh}}$). This discontinuity factor is moderate given that strong shocks could lead to enhancements by a factor of 4, and even up to 9 as has been shown for the case of Vela\,X-1 \citep{Gvaramadze2018}. Density and temperature jumps were also needed to explain the X-ray bow shock observed in Cygnus A \citep{Clarke1997}. Another extreme example of a single object showing spatially distinct regions due to the different shock regions is in the case of a high velocity pulsar B1957+20, where the X-ray and H$_\alpha$ emission trace the termination shock and the forward shock, respectively \citep{Stappers2003}. In the case of Cyg\,X-1, a smoking gun of this would be a clear spatial differentiation between the radio and optical emission regions, which is currently not possible due to the resolution limitations of both the radio and optical images of the bow shock structure. The presence of substructure, background/foreground point-sources also makes it complicated to compare the radio and optical bow shock maps. 

The density values of the unshocked ISM (6-7\,$cm^{-3}$) are consistent with the complex environment of Cyg\,X-1 given that there is a massive, bright, star-forming region namely the Tulip Nebula very close to Cyg\,X-1. Additionally, if the newly discovered extended emission of The Whale turns out to be at a similar distance as that of Cyg\,X-1, this would be another over-dense region close to Cyg\,X-1. The unshocked ISM density values are in agreement with high-density ISM surroundings of Vela\,X-1 \citep{Eijnden2022} and GRS\,1915+105 \citep{Motta2025}. Conversely, if the integrated surface brightness of the Cyg\,X-1 bow shock was at the same level as the rms around the region, we would not have been able to detect it in our observations. Thus, if the bow shock existed but was undetected in our L-band observations, it would imply that the electron density of the shocked ISM was between 1-3cm$^{-3}$, and the unshocked ISM would be $\approx$0.5cm$^{-3}$. This provides an upper limit of the ISM densities of the environments around other XRBs where such jet--ISM interaction sites have not yet been detected in the ThunderKAT/XKAT data.

\subsection{Velocity and age of the bow shock}\label{velocity}
Based on Rankine–Hugoniot conditions, which describe the parameters of the medium on both sides of a shock front, the expansion velocity ($\dot{L}$) of a strong shock (i.e. a very high Mach number) in a mono-atomic gas can be approximated as \citep{Landau1991, Kaiser2004}  
$$ \dot{L} \approx \sqrt{\frac{16 k_b}{3 m_p}T},$$
where $m_p$ is the mass of the proton and $T$ is the temperature of the gas. 
The velocity of the bow shock can be used to estimate the velocity of the leading edge of the jet that is driving into the ISM and creating the bow shock, since \citet{Kaiser1997} showed that these two velocities are roughly similar. Using this expression, we estimate the velocity of the bow shock to be $21\,$km\,s$^{-1}<\dot{L} <\,364\,$km\,s$^{-1}$ for a temperature range $10^4$K$\,< T <3\times10^6$K. This velocity is consistent with results from G05, as the same temperature range was used in both cases. \par
Since the expansion velocity of a bow shock driven by a jet is nearly equal to the velocity of the jet's driving end, the bow shock's age can provide constraints on the age of the jet itself \citep{Kaiser1997,Kaiser2004}. We adopted the methodology and assumptions of \citet{Kaiser2004}, who modelled the jet--ISM interaction sites near GRS\,1915+105 (also \citealt{Motta2025}). Both the approaches mentioned above use the basic foundation built in \citet{Kaiser1997} to model the formation and evolution of hotspots and bow shocks around FR{\sc{ii}} type galaxies and were later applied to the features around the BHXB GRS\,1915+105. The large-scale structures around FR{\sc{ii}} galaxies are approximated as self-similar features, a result attributed to an ambient density profile that decreases as a function of the square of the distance to the bow shock \citep{Scheuer1974}. \citet{Falle1991} updated this model by suggesting that the self-similar evolution of a cocoon needed the density profile of the ambient medium to be a function of $\frac{1}{d^2}$ or steeper. Finally, \citet{Kaiser1997} were able to show that both the cocoon and the bow shock would evolve showing self-similarity if the ISM density around the source was less steep than $\frac{1}{d^2}$. This helped them conclude that the length of the jet that powers the bow shock grows with time as $L_j \propto t^{3/5}$, assuming a uniform environment of the source. The above models are all approximate since the ambient density is not always as simple as a power law, nor are large-scale structures always due to the jets being self-similar. However, in the absence of a complete physical model, these approximations will provide a good starting model that can be applied to XRBs too. Hence, we assume that in our model (like for AGNs and in the case of \citet{Kaiser2004}), the lifetime of the large-scale structure of the jet can be estimated as a ratio of the length of the jet to the velocity of the apex of the bow shock and is given by

$$ t = \frac{3 L_j}{5 {{\dot{L}}_{j}}}. $$
Using the range of $\dot{L_j}$ derived in Sect. \ref{velocity} and the distance of the apex of the bow shock from the Cyg X-1 as $L_{j}$, we calculate the lifetime of the bow shock of Cyg\,X-1 to be in the range 0.04--0.3\,Myr, where the limits are the 16 and 84 percentiles of the resulting distribution. We assume that the duty cycle of the jets of Cyg\,X-1 is 90$\%$ based on the observation that the radio jets of Cyg\,X-1 are in a hard state and power collimated jets for 90$\%$ of its lifetime \citep{Gallo2005}. Thus, we can extrapolate this assumption in claiming that the lifetime of the large-scale structure and the creation of jets from Cyg\,X-1 is the same, and hence the age of the jets is estimated to be 0.04--0.3\,Myr. The jet that is responsible for the bow shock structure is orders of magnitude younger than the lifetime of Cyg\,X-1 itself, which is estimated to be between 5-7\,Myr \citep{Mirabel2003}. 

\subsection{Energy transport rate}
It has been shown for FR{\sc{ii}} radio galaxies that a constant supply of energy is required to sustain the radio cocoons and bow shocks around these AGNs On large scales, it is expected that over the lifetime of the jet, the energy transport rate of the jet is constant. \citet{Falle1991} suggested that the evolution of large-scale structure is self-similar beyond the characteristic length scale, and \citet{Kaiser2004} showed that this was indeed the case for XRBs like GRS\,1915+105. The characteristic length is calculated as
$$ L_0 = [\frac{Q_0^2}{\rho_0^2 c^6(\gamma_j -1)}]^{1/4},$$
where $Q_0$ is the energy transport rate, $\rho_0$ is the density of the unshocked ISM and $\gamma_j$ is the Lorentz factor of the jets. We assume a $\gamma_j$ of 1.1 for this calculation to maximise the characteristic length estimation. Here, we show that the characteristic scale length of the jet (8$\times$10$^{-5}$\,pc) is much smaller than the large scale jet length that is powering the bow shock in Cyg\,X-1, which suggests that the modelling used to estimate the time-averaged energy of the jets in GRS\,1915+105 applies to Cyg\,X-1. 

Thus, we can estimate the power of the jet ($Q_{jet}$) by using our calculations of the density of the ISM (n$_\mathrm{eS}$), the length of the jet ($L_j$), the velocity of the bow shock ($\dot{L_j}$) and a constant (C$_1$). C$_1$ depends on the opening angle and thermodynamical properties of the jet. The opening angle of the low-hard state jet of Cyg\,X-1 was inferred to be 0.3$^{\circ}$-- 2$^{\circ}$ using high-resolution imaging capabilities of the Very Long Baseline Array. However, the opening angle of the jet has not been directly measured in the intermediate X-ray state of Cyg\,X-1 and it is possible that the opening angle changes between different X-ray spectral states of a BHXB. Thus, here, we use a wider uniform distribution prior of the opening angle to be 0$^{\circ}$-30$^{\circ}$ for estimating the energy transport rate. The calculation and variation of C$_1$ with opening angle is presented in the accompanying Jupyter notebook. The equation used to estimate the energy transport rate is below for reference:
$$ Q_{jet} = (\frac{5}{3})^3~\frac{\rho_0}{C_{1}^5}L_j^2 ~\dot{L_j}^3.$$
We determine that the energy transport rate of the jet that powers the bow shock of Cyg\,X-1 is of the order of 2$\times$10$^{34}$\,ergs s${^{-1}}<$ Q$_{jet}^{b}<$ 1$\times$10$^{37}$\,ergs s$^{-1}$ for the opening angles 0$^{\circ}$-30$^{\circ}$. For the more stringent range of opening angles (0.3$^{\circ}$-- 2$^{\circ}$), the energy transport rate is in the range 2$\times10^{31}$\,ergs s$^{-1}<$ Q$_{jet}^{a}<$ 1$\times$10$^{35}$\,ergs s$^{-1}$.
The constraints on the energy transport rate of the jet averaged over the jet lifetime for Cyg\,X-1 is measured to be lower than that estimated for GRS\,1915+105. This is expected given that the ISM in which Cyg\,X-1 is embedded is less dense (6-7 particles cm$^{-3}$) than that of the region surrounding GRS\,1915+105 \citep[100 - 160 particles cm$^{-3}$;][]{Motta2025}, and the scale length of the jet is larger for GRS\,1915+105. To use this hierarchy of jet power to represent the true jet powers of the two XRBs, we need an estimate of the age of the XRBs. Cyg\,X-1 is a high mass XRB whose accretion disc is fed by the strong winds from its massive companion \citep{Cechura2015}, with an estimated age of 10$^{6}$\,yrs \citep{Mirabel2003}. On the other hand, in the case of GRS\,1915+105, the accretion is sustained by Roche lobe overflow, with the system being reported to have near-Eddington accretion for a large fraction of its active phases \citep{Done2004}, and it is also older than Cyg\,X-1 \citep{Podsiadlowski2003}. The estimations for jet power for both Cyg\,X-1 and GRS\,1915+105 are lower than the lower limit for the power distribution of jetted AGNs \citep[10$^{43}$\,ergs s$^{-1}$;][]{Fan2019}.

We can also estimate the pressure inside this bow shock and it is given by
$$ p = 0.0675 \frac{C_1^{10/3}}{R^2}(\frac{\rho_0 Q_0^2}{L_j^4})^{1/3}.$$
Thus, the pressure is estimated as 3$\times$10$^{-13}$\,ergs cm$^{-3}<$ p$<$ 2$\times$10$^{-11}$\,ergs cm$^{-3}$. Hence, the Cyg\,X-1 bubble has a lower pressure as compared to the GRS\,1915+105 lobe (1.9$\times$10$^{-11}$ -- 3$\times$10$^{-9}$ erg cm$^{-3}$).  

Most of the above analysis was based on the bow shock being a thermal emitter. However, some regions of the bow shock are also consistent with -1 spectral index, although the error bars are large in those regions. So, for completeness, we also estimated the minimum magnetic field associated with non-thermal emission of the bow shock. This minimum magnetic field occurs at close to the equipartition of energy between the electron population and the magnetic fields present in the emitting region. The equipartition magnetic field is given by 
$$B_{\rm eq} = \left(6 \pi \frac{\eta}{f} c_{12} \frac{L}{V}\right)^{2/7},$$
where $f$ is a filling factor (assumed 1), $c_{12}$ is a constant that relates the energy stored in the electrons to the magnetic field, $L$ is the total luminosity of the emitting region across the frequency band and $V$ is the volume of the emitting region. For an assumed spectral index of -0.9 informed by the uncorrected values of our spectral index map, the minimum magnetic field is $>$1$\times$10$^{-6}$\,G for L-band and $>$8$\times$10$^{-7}$\,G for S-band luminosity. These values are lower than those estimated for the bow shock around GRS\,1915+105 \citep{Motta2025}.

\subsection{Bow shock morphology}
The velocity of bow shock expansion can also be measured directly in the plane of the sky by measuring the change in the position of the bow shock over time. However, such measurements of faint diffuse emission like that of the Cyg\,X-1 bow shock would require either an extremely long time baseline or a very fast bow shock. As a test, we compared the bow shock location and morphology in G05 and the bow shock location as measured by our observations (see Fig. \ref{fig:zooms_Lband}, right panel). The bow shock observations although separated by 17 years, are almost aligned with each other as is seen by the contours of the L-band MeerKAT bow shock overlaid on the G05 image of the bow shock with WSRT. There is some apparent change in the morphology of the bow shock. The leading edge of the bow shock is seen connected leading to a complete U-shaped arc in G05. On the other hand, the leading edge in the most recent L-band observations with MeerKAT shows that the edge is as faint as the noise in the image with some connecting structure visible, which is at a 2$\sigma$ level. However, the leading edge of the bow shock that is barely detected in the L band is significantly detected in the S band. A longer observation of the bow shock in the L band will help improve the $uv$ coverage and also increase the sensitivity, aiding in assessing whether the leading edge of the bow shock persists today in the L band. \par 
As a direct consequence of the S-band MeerKAT image being more sensitive than the L-band image, we can identify new features in the bow shock, like the two smaller arcs (marked as B4 and B5 in Fig. \ref{fig:zooms_Sband}, middle panel, and seen clearly in the Fig. \ref{fig:zooms_Sband}, left panel) within the larger bow shock. Hints of these smaller structures are also seen in the L-band MeerKAT image, but are confusion-limited. We have overlaid the contours of the L-band bow shock on the S-band image in Fig. \ref{fig:zooms_Sband} (right panel). These smaller arcs were seen as a solid extended structure between the arms of the U-shaped arc of the bow shock near the leading edge in G05. Here, we present three tentative interpretations of the detection of these smaller arcs. The first one is that the smaller arcs are the result of different transient jet-ejection episodes that might have occurred over the lifetime of Cyg\,X-1. This would indicate that some jets have been launched misaligned from the usual, perpendicular jet axis to the accretion disc. These misaligned jets end up sweeping different, smaller regions as compared to the more stable, compact jets that are responsible for the larger U-shaped structure. A second possibility is that depending on the relative location of the black hole and the massive companion star, the stellar winds can bend the extended jet even when it is in the low-hard state creating the smaller arcs \citep{Bosch-Ramon2016}, and at larger distances from these smaller arcs, the shocked front could be sweeping up gas and creating an even larger interference shock front pattern. The third possibility is that the smaller arcs are seen due to projection effects, whereby small under (or over) densities in the local ISM could enable the shocks to break out of the overall structure. The two-dimensional image shown in Fig. \ref{fig:zooms_Sband} is a three-dimensional structure, so these smaller arcs may be localised in regions that are seen in projection on the overall bow shock structure. 
However, we note that the scenarios presented above are currently only hypotheses as the smaller arcs are too faint to be able to do detailed modelling. \par 
\subsection{The Whale}
Other than the main subject of these observations, the Cyg\,X-1 bow shock, there is another noticeable large-scale structure in Figs. \ref{fig:full_map_Lband} and \ref{fig:full_map_S_band}. The combined emission of W1 and W2 appears to be shaped like a whale and hence we have named the emission structure \lq The Whale\rq . We added annotations in Figs. \ref{fig:full_map_Lband} and \ref{fig:full_map_S_band} to showcase the structure. Currently, there are no distance estimates for these emission regions; thus, we only present a qualitative discussion of these large-scale structures, as multi-wavelength follow-up and deeper radio observations of this region are needed to understand the nature and driving force of this emission. The Whale could be a single large diffuse structure, or two independent regions that appear to be connected here due to the limited dynamic range of the image. The tail of The Whale (W1) could be a shock interface or a H{\sc{ii}} region and has an average brightness of 109\,\uJy in L band and 40\,\uJy in the S band. We report a bright (6\,mJy in L band and 11\,mJy in the S band), compact object at a position of RA $19^{h}:57^{m}:35^{s}.687$ and Decl $35^{d}:23^{m}:38^{s}.374$ and at an angular distance of 3\,arcmin from the apex of the arc-like structure, W1. We used Vizier\footnote{https://vizier.u-strasbg.fr/viz-bin/VizieR} to check if this source had any optical counterparts, and there were three \textit{Gaia} matches within the 7\,arcsec radius of this radio source. Two of these \textit{Gaia} sources have negative parallax, whereas one had a parallax of 0.3$\pm$0.2, which hints at its possible Galactic nature. There are three more radio sources near W1 on the same side of this radio source, but none of them had compelling \textit{Gaia} counterparts. We will conduct a detailed analysis of the possible association and interaction site W1 in a follow-up study.   

\section{Conclusions}\label{Section 6}
Jet--ISM interaction sites are a promising laboratory for estimating the energetics of the jet powering the site. In this study we extrapolated models developed for AGNs to the BHXB Cyg\,X-1. We observed the Cyg\,X-1 region using the MeerKAT telescope in two frequency bands, the L band (central frequency 1.28\,GHz) and the S band (central frequency 2.64\,GHz). We detected the jet-powered bow shock as a U-shaped arc in both frequencies, with an average brightness of 110$\pm$24\,$\mu$Jy\,bm$^{-1}$ and 38$\pm$13\,$\mu$Jy\,bm$^{-1}$ in the L and S bands, respectively. We were able to resolve smaller arcs and finer bow shock structure in the S-band image, which shows that the morphology of the bow shock is more complicated than a single U-shaped edge brightened bow shock structure. We report the spectral index map of the bow shock and find that the distribution of spectral indices is uniform throughout the structure. In the future we will conduct observations in other frequency bands to accurately measure the absolute spectral index. \par
We combined the observed bow shock parameters from our L- and S-band observations to estimate the density of unshocked ISM (6-7\,cm$^{-3}$) within a temperature range (1$\times$10$^4$-3$\times$10$^6$\,K) that could explain the emissivity in the radio bands. We modelled the bow shock around Cyg\,X-1 and determined the age of the Cyg\,X-1 jets to be 0.04--0.3\,Myr and the power of the jet feeding the bow shock structure to be 2$\times10^{31}$\,ergs s$^{-1}<$ Q$_{jet}^{a}<$ 1.0$\times$10$^{35}$\,ergs s$^{-1}$ for the case of opening angles of 0.3$^\circ$--2.0$^\circ$. This well-constrained jet power is lower than the earlier estimate for this bow shock and also lower than the only other BHXB for which this has been determined, GRS\,1915+105. The success of these observations suggests that a systematic search for bow shocks around XRBs is a great way to advance studies of black hole jet powers, outputs, and jet--ISM interaction and densities, and to tie inflows and outflows of accreting black hole systems. The future generation of radio telescopes like SKA, SKA-mid, and ngVLA will be able to make drastic advancements as they will be able to probe bow shocks even in lower-ISM-density environments.

\begin{acknowledgements}
PA is supported by the WISE fellowship program, which is financed by NWO. SEM acknowledges support from the INAF Fundamental Research Grant (2022) EJECTA. 
JvdE acknowledges a Warwick Astrophysics prize post-doctoral fellowship made possible thanks to a generous philanthropic donation, and was supported by funding from the European Union's Horizon Europe research and innovation programme under the Marie Skłodowska-Curie grant agreement No 101148693 (MeerSHOCKS) for part of this work.  \\
ASTRON, the Netherlands Institute for Radio Astronomy, is an institute of the Dutch Research Council (De Nederlandse Organisatie voor Wetenschappelijk Onderzoek, NWO). We acknowledge the use of the Ilifu cloud computing facility – www.ilifu.ac.za, a partnership between the University of Cape Town, the University of the Western Cape, Stellenbosch University, Sol Plaatje University and the Cape Peninsula University of Technology. The Ilifu facility is supported by contributions from the Inter-University Institute for Data Intensive Astronomy (IDIA – a partnership between the University of Cape Town, the University of Pretoria and the University of the Western Cape), the Computational Biology division at UCT and the Data Intensive Research Initiative of South Africa (DIRISA).\\
The authors are grateful to Elena Gallo for providing the archival radio map of the Cyg\,X-1 bow shock. The authors would like to thank Fraser Cowie, David Russell, Payaswini Saikia for their inputs on the manuscript of the paper, and Andrew Hughes and Joe Bright for scheduling the MeerKAT observations. 
\end{acknowledgements}

\section*{Data availability}\label{Section 8}
The datasets used in this paper are available in the public archive of SARAO at \url{https://archive.sarao.ac.za/}. The code used to conduct the analysis, create plots and maps is available here \url{https://github.com/pikkyatri/CygX-1-bow-shock}.

\bibliographystyle{aa} 
\bibliography{aa52837-24} 

\begin{appendix}
\section{Additional figure}

\begin{figure}[h!]\centering
\includegraphics[width=1.0\textwidth]{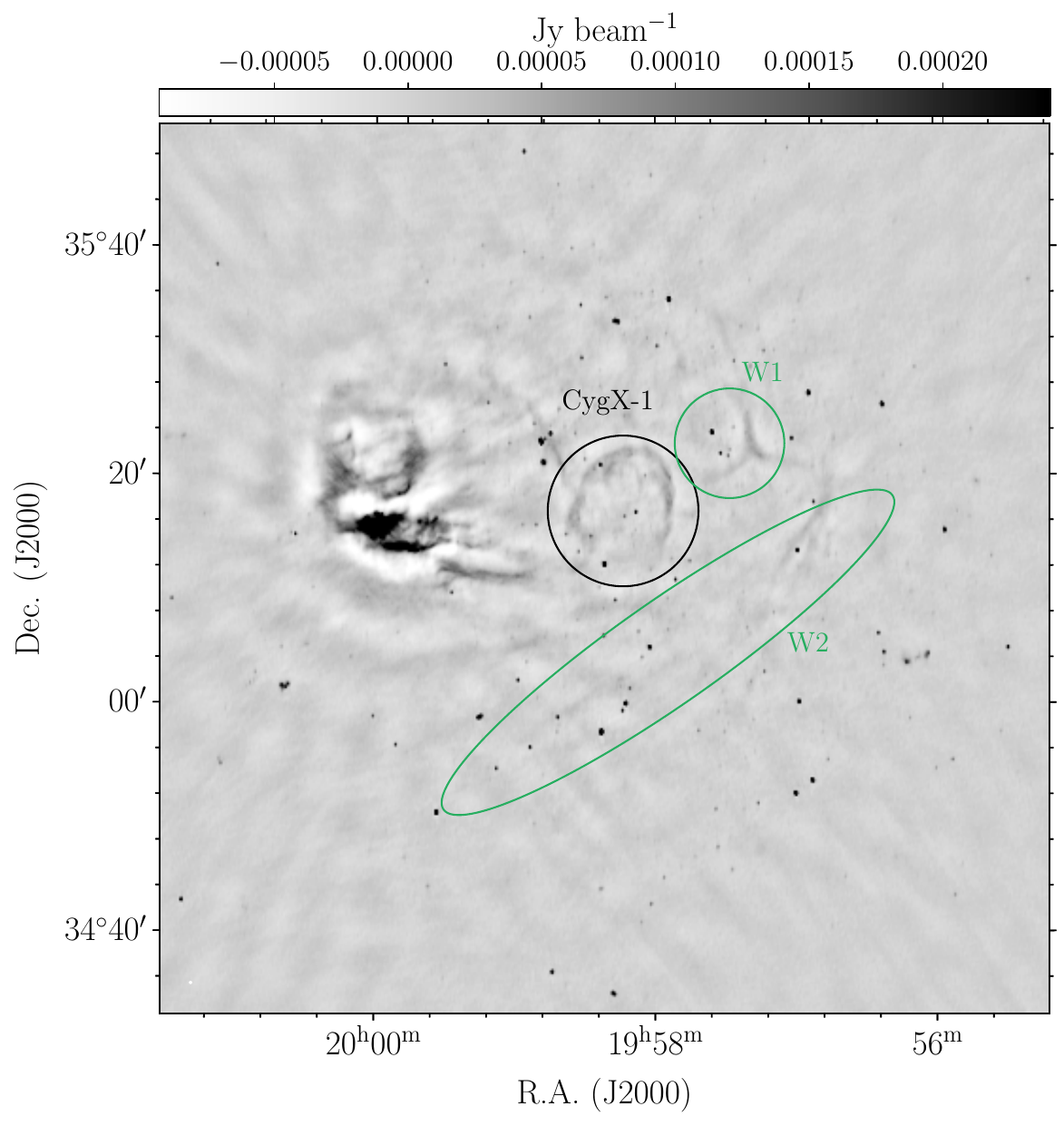}
\caption{Cyg\,X-1 field as observed by the MeerKAT telescope in the S band after applying a minimum $uv$ cut of 500 wavelengths. A black circle marks the location of Cyg\,X-1 and the bow shock. The locations of W1 and W2 as detected in the L-band image are marked in green. The W2 has become fainter, indicating that this emission was being picked up by the short baselines that have been flagged out here but are present in the L-band image.}
\label{fig:full_map_S_band_uv_cut} 
\end{figure}

\end{appendix}
\end{document}